\documentclass[notitlepage,nofootinbib,preprintnumbers,aps,prd]{revtex4-1}
\usepackage{amsmath,amssymb,natbib,bm,color}
\usepackage{appendix}
\usepackage{graphicx}
\usepackage{hyperref}
\usepackage{slashed}
\newcommand{\be}{\begin{equation}}
\newcommand{\ee}{\end{equation}}
\newcommand{\bea}{\begin{eqnarray}}
\newcommand{\eea}{\end{eqnarray}}

\begin{document}

\title{On the Anisotropy of the Stochastic Gravitational Wave Background\\ from Sub-Horizon-Collapsed Primordial Black Hole Mergers}
\author{Stefano Profumo$^{1,2}$}
\author{Fengwei Yang$^3$}

\affiliation{$^1$Department of Physics, University of California, Santa Cruz
	Santa Cruz, CA 95064, USA
}
\affiliation{$^2$Santa Cruz Institute for Particle Physics,
	Santa Cruz, CA 95064, USA
}

\affiliation{$^3$Department of Physics and Astronomy, University of Utah, Salt Lake City, UT 84112, USA}

\begin{abstract}
We study the properties of the stochastic gravitational wave background (SGWB) resulting from the mergers of primordial black holes (PBH) that formed from the collapse of sub-horizon regions in the early universe. We adopt a model-independent approach, where we parameterize the fraction $f_H$ of the wavelength of the perturbation mode in units of the horizon radius when the patch starts to gravitationally collapse. Assuming a monochromatic spectrum of isocurvature perturbations and spherically-symmetric density perturbations, we investigate the isotropic SGWB energy density and angular power spectrum at various frequencies, PBH masses, and horizon size fractions. The key effect of sub-horizon formation is a change in the PBH mass function and formation redshift, which, in turn, affects gravitational wave (GW) observables. We find that sub-horizon PBH formation in general {\em enhances} the isotropic SGWB energy density and the absolute angular power spectrum. However, the quasi-monotonic increases in both quantities as $f_H$ decreases cease when the chirp mass of the binary PBHs reaches a mass threshold determined by the frequency of observation; the isotropic SGWB energy density spectrum significantly drops above the corresponding cutoff frequency. 
\end{abstract}

\maketitle

\section{Introduction}





The intriguing possibility that black holes form from large density fluctuations in the early universe instead of from the gravitational collapse of large astrophysical objects in the late universe was proposed long ago \cite{Hawking:1971ei}. Such ``primordial black holes'' (PBHs) might be related to the cosmological dark matter \cite{Carr:2020xqk}, to the matter-antimatter asymmetry (see e.g. \cite{Smyth:2021lkn} and references therein), or both \cite{Morrison:2018xla}; they could play a role in seeding supermassive black holes \cite{Bean:2002kx}, or even explain part (or perhaps all) of the black hole-black hole mergers observed by gravitational wave (GW) interferometers \cite{Bird:2016dcv}. While the direct detection of even a single black hole with a mass below the Tolman-Oppenheimer-Volkoff limit \cite{Belczynski:2009xy} would indicate the existence of PBHs (or of unexpected new physics \cite{Giffin:2021kgb}), other observables may point towards a non-astrophysical origin for black holes. For instance, the spin distribution expected from PBHs, whether produced in matter- or radiation-domination in the early universe, is expected to be markedly, or at least statistically significantly different than that expected for astrophysical black holes \cite{Fernandez:2019kyb}.

While no evidence for a stochastic background of GWs has been conclusively detected yet (see however \cite{NANOGrav:2020bcs}), there are several reasons to believe that it will soon be. A perfectly plausible source of such diffuse background are unresolved binary black hole mergers at all redshifts, including of PBHs \cite{Shannon:2015ect}. Some information on the origin and production mechanism of the black holes whose binary mergers might contribute to the diffuse background of GWs may arise from spectral considerations \cite{Raidal:2017mfl}. Here, we rather focus on the use of the power spectrum of anisotropies of the stochastic background as a possible tell-tale signature of the production mechanism of the black holes (for previous studies see e.g. \cite{Bartolo:2019zvb} and references therein).

Previous studies that focused on utilizing the anisotropy of the stochastic gravitational wave background (SGWB) as a metric to differentiate PBHs from astrophysical black holes include, for instance, Ref. \cite{Wang:2021djr} (see also Ref. \cite{Bavera:2021wmw} on how to utilize the isotropic SGWB to constrain the primordial origin of BH formation channel). Here, however, we take a different perspective, and aim at differentiating one particular aspect of PBH formation: the portion of the Hubble horizon that collapsed at formation. Generically, PBHs result from the collapse of patches of the early universe as large as the Hubble horizon at collapse, but depending on the formation mechanism the collapsing region can be significantly smaller. We therefore simply assume, here, in a model-agnostic way, that PBHs result from the collapse of a fraction $f_H<1$ of the Hubble horizon only (we discuss expectations for $f_H$ in different scenarios below), and study the impact of $f_H$ on the anisotropy of the SGWB.

The remainder of this study is as follows: in the next section we discuss PBH formation from the collapse of curvature perturbations in the standard and in the presently investigated non-standard scenarios. Sec.~\ref{sec:aniso} describes the calculation of the power spectrum of anisotropies of the SGWB, and presents our results. The final Sec.~\ref{sec:discussion} presents our discussion and conclusions.

\section{PBH formation from curvature perturbations}
 PBHs can  form in the early universe as a result of the gravitational collapse of curvature perturbations $\zeta(x)$. The power spectrum of  curvature perturbations, $P_\zeta(k)$, determines the width of the Gaussian distribution of matter overdensities $\delta\rho(x)/\rho(x)\equiv\delta(x)$. The distribution of matter overdensities then in turn determines the mass function of the resulting PBHs. According to the standard scenario, where a PBH is formed at the horizon re-entry time $t_{ H}\sim 1/k_*$, with $k_*$ the scale of the perturbation that collapses to form the PBH, the mass of the PBH is directly related, and proportional to, the Hubble mass -- the mass enclosed at that time in a Hubble patch. The later the scale $k_*$ reenters the horizon, the larger the PBH mass. Notably, and generically, however, {\em PBHs can be also formed at the sub-horizon scales}, and the formation time can be {\em much later} than $t_H$, modifying the relation between PBH mass and horizon mass. Here, we explore the implications of sub-horizon-forming PBHs for the anisotropy of the gravitational wave spectrum induced by the PBH merger.

\subsection{The standard scenario for PBH formation}
The horizon reentry time is defined by, in natural units,
\be
\label{reentry}
t_{ H}\equiv \frac{1}{k_*},
\ee
where $k_*$ is the scale of the over-density that will collapse to form a PBH, and where we denote the characteristic wavelength of the density perturbation as $\lambda_*\equiv 1/k_*$. The size of a Hubble patch at horizon re-entry  is given by 
\be
r_{H}=t_{H}.
\ee
Thus, we can use Eq.~(\ref{reentry}) to relate the size of the Hubble patch at the horizon reentry time and the characteristic wavelength of the density perturbation,
\be
\lambda_* = r_{ H}.
\ee

Statistically-rare perturbations (such that one can neglect hierarchical random fields) are expected to be approximately spherically symmetric \cite{Bardeen:1985tr}. With the assumption that the perturbation is spherically symmetric, the amplitude of the smoothed density contrast $\bar{\delta}$ is related to the curvature perturbation, in radiation domination, as
\be
\label{eq:delta1-def}
\bar{\delta}=-\frac{2}{3}r_m\zeta'(r_m)(2+r_m\zeta'(r_m))=\delta_1-\frac{3}{8}\delta_1^2,
\ee
where $r_m$ is the smoothing scale, the prime is the derivative with respect to the radial coordinate, and $\delta_1$ is the linear component of the $\bar{\delta}$, given by
\be
\delta_1=-\frac{4}{3}r_m\zeta'(r_m).
\ee
The distribution of $\delta_1$ obeys a Gaussian probability distribution function (PDF):
\be
{\rm PDF}(\delta_1)=\frac{1}{\sqrt{2\pi \sigma^2}}\exp\left(-\frac{\delta_1^2}{2\sigma_0^2}\right),
\ee
where the variance $\sigma_0^2$ characterizes the width of the Gaussian PDF, and it is obtained from the zeroth moment of the power spectrum of $\delta_1$.
The $j$-th moments of the power spectrum $P_{\delta_1}$ are given by
\be
\sigma_j^2=\int_0^\infty \frac{dk}{k}{P}_{\delta_1}(k,r_m)\left(\frac{k}{aH}\right)^{2j}.
\ee
The variance is 
\be
\label{sigma2}
\sigma^2_0=\langle \delta_1^2\rangle=\int_0^\infty \frac{dk}{k}{P}_{\delta_1}(k,r_m)=\frac{16}{81}\int_0^\infty \frac{dk}{k} (kr_m)^4 \tilde{W}^2(k,r_m)T^2(k,r_m)P_\zeta(k),
\ee
where we choose a Dirac-$\delta$ peak power spectrum of the curvature perturbation $P_\zeta(k)=A_s\delta\left(\ln\frac{k}{k_*}\right)+A_b~(A_s \gg A_b)$, $A_s$ and $A_b$ are the amplitudes of the short-wavelength mode and background mode, respectively, $r_m$ is the smoothing scale that determines how to choose the spatial volume to average the overdensity peak, 
\begin{equation}\tilde{W}(k,r_m)=3\frac{\sin(kr_m)-kr_m\cos(kr_m)}{(kr_m)^3}
\end{equation} is the Fourier transform of the top-hat window function with the time-dependent smoothing scale $r_m$, and 
\begin{equation}T(k,r_m)=3\frac{\sin(kr_m/\sqrt{3})-kr_m/\sqrt{3}\cos(kr_m/\sqrt{3})}{(kr_m/\sqrt{3})^3}
\end{equation} is the linear transfer function. In the literature \cite{Young:2019yug}, it is common to choose the smoothing scale to be the Hubble scale, i.e., $r_m(z)=r_H=(aH)^{-1}$.

When the smoothed density contrast $\bar{\delta}$ is greater than the critical threshold value $\delta_c$, a PBH is formed, with a mass  closely related to the modeling of the gravitational collapse,
\be
\label{eq:mpbh}
M_{\rm PBH}=\mathcal{K}(\bar{\delta}-\delta_c)^\gamma M_{ H},
\ee
where $M_H$ is the horizon mass at the formation time of PBH, $\mathcal{K}=4$ accounts for most of the shape of the density contract, $\gamma=0.36$ is the critical exponent dependent on the equation of state of the universe and here is evaluated in the radiation-dominated era \cite{Young:2019yug}, and $\delta_c=0.51$ is chosen for a typical profile shape of the density perturbation and a real-space top-hat window function at the horizon reentry time \cite{Musco:2018rwt,Young:2019osy}. Given the above PBH formation model, the distribution of the Gaussian random field $\delta_1$ thus determines the abundance and mass function of the resulting PBHs. By inverting Eq.~(\ref{eq:mpbh}) and combining with Eq.~(\ref{eq:delta1-def}), one can obtain the value of $\delta_1$ corresponding to a given PBH mass $M_{\rm PBH}$, 
\be
\delta_1(M_{\rm PBH})=\frac{2}{3}\left(2-\sqrt{4-6\delta_c-6\left(\frac{M_{\rm PBH}}{\mathcal{K}M_H}\right)^{1/\gamma}}\right).
\ee

The initial abundance of PBHs is usually described by the energy density fraction of PBHs in the universe at a single formation time,
\be
\label{eq:beta}
\beta=\int_{\mu_{c}}^{\mu_{\rm max}}d\mu \frac{M_{\rm PBH}(\mu)}{M_{ H}}n_{\rm peak}(\mu),
\ee
where $\mu\equiv\delta_1/\sigma_0$, $\mu\in[\mu_c,\mu_{\rm max}]$, $\mu_c=\delta_{c,1}/\sigma_0$, $\mu_{\rm max}=\delta_{1,{\rm max}}/\sigma_0=4/(3\sigma_0)$, makes the gravitational collapse happen,  
$\delta_{c,1}=\frac{4}{3}\left(1-\sqrt{1-\frac{3}{2}\delta_c}\right)$ is the critical value for gravitational collapse of $\delta_1$, $n_{\rm peak}=\frac{1}{4\pi^2}\left(\frac{\sigma_1}{\sigma_0}\right)^3\mu^3\exp\left(-\frac{\mu^2}{2}\right)$ is the number density of over-density peaks in a comoving volume \cite{Bardeen:1985tr}.

The mass function is defined as 
\be
\Psi(M_{\rm PBH})=\frac{1}{f_{\rm PBH}}\frac{d f_{\rm PBH}}{dM_{\rm PBH}},
\ee
such that $\int\Psi(M_{\rm PBH})dM_{\rm PBH}=1$, where $f_{\rm PBH}\equiv\Omega_{\rm PBH}/\Omega_{\rm DM}$ is the fraction of the present dark matter (DM) energy density stored in PBHs. The fraction $f_{\rm PBH}$ can be calculated from $\beta$ by integrating the contributions from PBHs formed across the history of the universe until the time of matter-radiation equality,
\be
\label{eq:fpbh-integral}
f_{\rm PBH}=\frac{1}{\Omega_{\rm DM}}\int_{M_{H,{\rm min}}}^{M_{H,{\rm max}}}d(\ln M_H)\left(\frac{M_{H,{\rm eq}}}{M_{H}}\right)^{1/2}\beta(M_H),
\ee
where $M_{H,{\rm eq}}=\frac{4\pi}{3}2\rho_{\rm eq}H_{\rm eq}^{-3}\approx2.8\times10^{17}M_\odot$ is the approximate horizon mass at the matter-radiation equality.
Assuming a Dirac-$\delta$ form for the power spectrum of the curvature perturbation and assuming as a result that all PBHs form at a single time, the mass function is given by
\be
\Psi(M_{\rm PBH})=\frac{1}{\beta}\frac{d\beta}{dM_{\rm PBH}}.
\ee
Here is the derivation. First, we relate $\beta$, the abundance of PBHs at a single formation time, to $\Omega_{\rm PBH}$, the energy density parameter of PBHs observed today: 
\be
\beta=\frac{\rho_{\rm PBH}}{\rho_r}=\frac{\rho_{\rm PBH}}{\rho_{\rm c,0}}\frac{\rho_{\rm c,0}}{\rho_{ r}}=\frac{\rho_{\rm PBH,0}\left(\frac{a_0}{a}\right)^3}{\rho_{\rm c,0}}\frac{\rho_{\rm c,0}}{\rho_{ r,0}\left(\frac{a_0}{a}\right)^4\left(\frac{g_0}{g}\right)^{1/3}}=\frac{\Omega_{\rm PBH}}{\Omega_r}\left(\frac{a}{a_0}\right)\left(\frac{g}{g_0}\right)^{1/3},
\ee
where $\Omega_r$ is the energy density parameter of radiation observed today, $a$ is the scalar factor, $a_0=1$ is the scalar factor today, $g$ and $g_0$ are the degree of freedom of the relativistic species at the time of $a$ and $a_0$ respectively. Therefore, the energy density fraction of PBHs in terms of DM is given by a simplified expression instead of the integral in Eq.~(\ref{eq:fpbh-integral}),
\be
f_{\rm PBH}=\frac{\Omega_{\rm PBH}}{\Omega_{\rm DM}}=\frac{\Omega_r}{\Omega_{\rm DM}}\frac{1}{a}\left(\frac{g_0}{g}\right)^{1/3}\beta.
\ee


Let $\psi(x)=\Psi\times M_{ H}$ be a dimensionless mass function:
\be\label{eq:massfunction}
\psi(x)=\frac{1}{4\pi^2\beta\sigma_0}x\left(\frac{\sigma_1}{\sigma_0}\right)^3\left(\frac{\delta_1(x)}{\sigma_0}\right)^3\exp\left(-\frac{\delta_1^2(x)}{2\sigma_0^2}\right)\frac{\sqrt{2}(x/\mathcal{K})^{1/\gamma-1}}{\gamma\mathcal{K}\sqrt{2-3\delta_c-3(x/\mathcal{K})^{1/\gamma}}},
\ee
where $x\equiv\frac{M_{\rm PBH}}{M_{ H}}$ is the ratio between PBH mass and horizon mass. The normalization of the dimensionless mass function becomes $\int\psi(x)dx=1$, and the PBH-mass-to-horizon-mass ratio $x$ ranges from 0 to 2.05, where the upper limit is coming from the that Eq.~(\ref{eq:delta1-def}) has a maximum value $\bar{\delta}=\frac{2}{3}$ such that 
\be
M_{\rm PBH,max}=\mathcal{K}\left(\frac{2}{3}-\delta_c\right)^\gamma M_{ H}=2.05M_{ H},
\ee
where the second identity is obtained by substituting the parameters of the PBH formation model used in this paper.

\subsection{Sub-horizon PBH formation}
Generically, a PBH can form from the collapse of a density perturbation of  $\lambda_*$ smaller than the size of the horizon by a factor $f_H$,
\be
\lambda_* = f_H \times r_H',
\ee
where we use the symbol $r_H'$ to indicate the Hubble patch where the PBH is formed. For example, Ref. \cite{Nakama:2016gzw} considers $f_H=1/4$. Several example formation mechanisms with $f_H\lesssim 1$ and/or with $f_H\ll 1$ have been proposed and discussed in detail in the literature. For instance, Ref.~\cite{Cotner:2019ykd} considers PBH production from the collapse of scalar field lumps (Q-balls or oscillons) that can potentially temporarily dominate the energy density of the universe leading to efficient PBH formation; the typical size of the lumps (formed, here, not because of gravitational instability, but rather due to self-interactions) is predicted, from numerical simulations, to be a fraction $f_Q\sim 10^{-2}-10^{-1}$ of the horizon size. Bubble collision upon spontaneous symmetry breaking and a first-order phase transition also generically seed sub-horizon density perturbations leading to collapse into black holes with mass much smaller than the horizon mass (see e.g. \cite{PhysRevD.26.2681}). Similarly, cosmic string loops lead to sub-horizon mass PBH, depending on the string tension \cite{PhysRevD.47.3265, PhysRevD.53.3002}. Ref.~\cite{Meissner:2020rzo} entertains a more speculative possibility of multigravitino Bogomol'nyi-Prasad-Sommerfield (BPS) states that if sufficiently large could collapse into black holes whose mass is entirely unrelated to the horizon mass. A somewhat less speculative possibility is the collapse of Fermi balls consisting of massive fermions trapped into regions of false vacuum in a first-order phase transition \cite{Kawana:2021tde}. Other similar possibilities include the formation of PBH from long-range scalar forces and scalar radiative cooling \cite{Flores:2020drq}, and a late-phase transition in a strongly interacting fermion-scalar fluid \cite{Chakraborty:2022mwu}. Finally, the collapse of neutron stars in the very late universe, possibly triggered by the accumulation of dark matter in the neutron star core, can also lead to the formation of black holes with masses much smaller than the horizon mass \cite{Acevedo:2020gro,Giffin:2021kgb}. 

For $f_H<1$, one has $r_H'= r_H/f_H>r_H$, i.e., the Hubble patch where the PBH is formed is greater than the one in the standard scenario, and thus the formation time is later than the one in the standard scenario, $t_H'>t_H$. Assuming PBHs formed in the early universe, during radiation domination, the formation redshift can be related to the one in the standard scenario by the following equation:
\be
\frac{z'_{\text {form}}} {z_{\text{form} } } = f_H.
\ee

\begin{figure}[!t]
\centering  
  \includegraphics[width=0.6\textwidth]{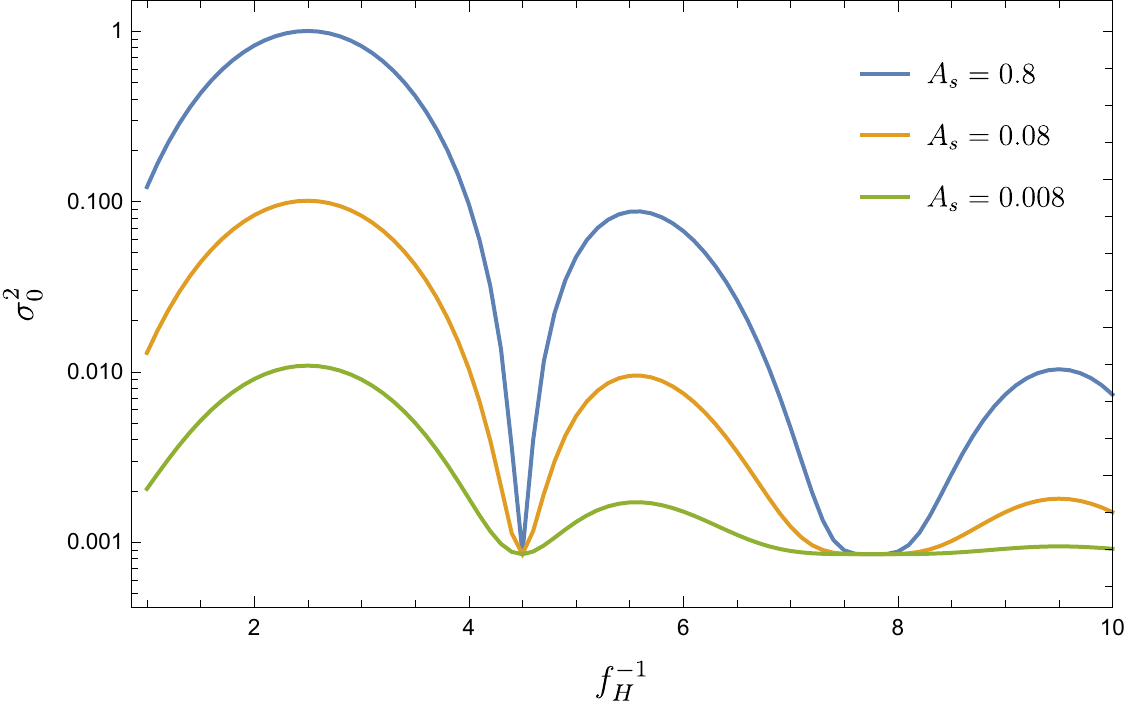}
 \caption{\label{fig:sigma2Ofn} The variance $\sigma^2_0$ of the over-density distribution as a function of $1/f_H$, for different amplitudes of the short-wavelength perturbation mode $A_s=0.8\,(\text{blue})$, $0.08\,(\text{orange})$, $0.008\,(\text{green})$, at a fixed value for the amplitude of the background mode $A_b=8\times10^{-4}$. 
 }
\end{figure}

 In the sub-horizon formation scenario,  $r_m(z)$ in Eq.~(\ref{sigma2}) is evaluated at time $z'_{\text {form}}$ instead of $z_{\text {form}}$, such that $k_* r_m(=f_H^{-1})\neq1$. There is one subtlety for the smoothing scale of the top-hat window function. In the standard case, there is only one density contrast peak in one Hubble patch, so choosing $r_m=r_H$ is reasonable. However, in the sub-horizon formation case, there are $\lfloor f_H^{-1}\rfloor$ density contrast peaks, but we still choose $r_m=r_H$ as the smoothing scale. We comment later on the impact of choosing different smoothing radii and how our results depend on the choice of $r_m$. 

Figure \ref{fig:sigma2Ofn} shows $\sigma^2_0$ as a function of the inverse of the horizon size fraction $f_H^{-1}$ for different values of $A_s$ with a fixed value of $A_b=8\times10^{-4}$. For a given $A_s$, $\sigma_0^2$ increases monotonically before reaching its maximum at $f_H^{-1}\sim2.5$, and it rapidly drops until it reaches its minimum at $f_H^{-1}\sim4.5$. The behavior of $\sigma_0^2$ is quasi-periodic with respect to $f_H^{-1}$ because the window function and transfer function in Eq.~(\ref{sigma2}), containing $\sin(k_*r_m)=\sin(f_H^{-1})$ and $\cos(k_*r_m)=\cos(f_H^{-1})$, are periodic but with a suppressed amplitude at larger $f_H^{-1}$. Figure \ref{fig:sigma2Ofn} additionally illustrates how the value of $\sigma_0^2$ decreases globally for a smaller value of $A_s$, because $\sigma_0^2$ characterizes the auto-correlation of the overdensity whose amplitude is determined by the amplitude of the curvature perturbation. We stress that the non-linear dependence of the value of $\sigma_0^2$ on $A_s$ at some values of $f_H$ stems from the fact that the ratio $A_s/A_b$ is not a constant. If $A_s/A_b$ were a constant, $\sigma_0^2$ would linearly depend on $A_s$, which can be directly derived from Eq.~(\ref{sigma2}). Starting from the Sec. \ref{sec:aniso}, for simplicity, we choose $A_b=0$, and the $j$-th order moments are simplified to be
\be
\label{eq:sigma_As_fH}
\sigma_j^2=432A_sf_H^{8-2j}\left(\sin\left(f_H^{-1}\right)-f_H^{-1}\cos\left(f_H^{-1}\right)\right)^2\left(\sin\left(f_H^{-1}/\sqrt{3}\right)-f_H^{-1}/\sqrt{3}\cos\left(f_H^{-1}/\sqrt{3}\right)\right)^2.
\ee
We note that the above dependence on $A_s$ and $f_H$ brings a degeneracy of $A_s$ when fixing $f_{\rm PBH}$ and $f_H$. This is because $f_{\rm PBH}$ is proportional to $\beta$, which, in turn, is only determined by $\sigma_0^2$ and $\sigma_1^2$ (see Eq.~(\ref{eq:beta})); specifically, $\beta$ has an exponential dependence on $\sigma_0^2$, and a power-law dependence on $\sigma_1^2$. This results in  $f_{\rm PBH}$ being approximately fixed by a given choice of $\sigma_0^2$. However, Eq.~(\ref{eq:sigma_As_fH}) can have multiple roots for $A_s$ when fixing $\sigma_0^2$ and $f_H$, which can also be appreciated by imagining a horizontal line in Figure~\ref{fig:sigma2Ofn} and noting how it would intersect with curves of different colors from a large $\sigma_0^2$ value to a smaller value. Therefore, $A_s$ has multiple values for given $f_{\rm PBH}$ and $f_H$. Also, due to the quasi-periodic behavior of $\sigma_0^2$ versus $f_H$, one has the freedom to pick $A_s$ such that $f_{\rm PBH}$ is sufficiently small at a given $f_H$. In the calculation of Sec. \ref{sec:aniso}, we fix $f_{\rm PBH}=0.01$ and choose the {\em smallest value of $A_s$ for each $f_H$}. Eq.~(\ref{eq:sigma_As_fH}) also manifests choosing different smoothing scales $r_m$ will change the value of $\sigma_0^{2}$ and $\sigma_1^2$. For the choice of $r_m=r_H$, $k_*r_m=f_H^{-1}$, while in general, one can choose $f_Hr_H\le r_m\le r_H$, i.e., $1\le k_*r_m\le f_H^{-1}$. 

\begin{figure}[!t]
\centering  
  \includegraphics[width=0.45\textwidth]{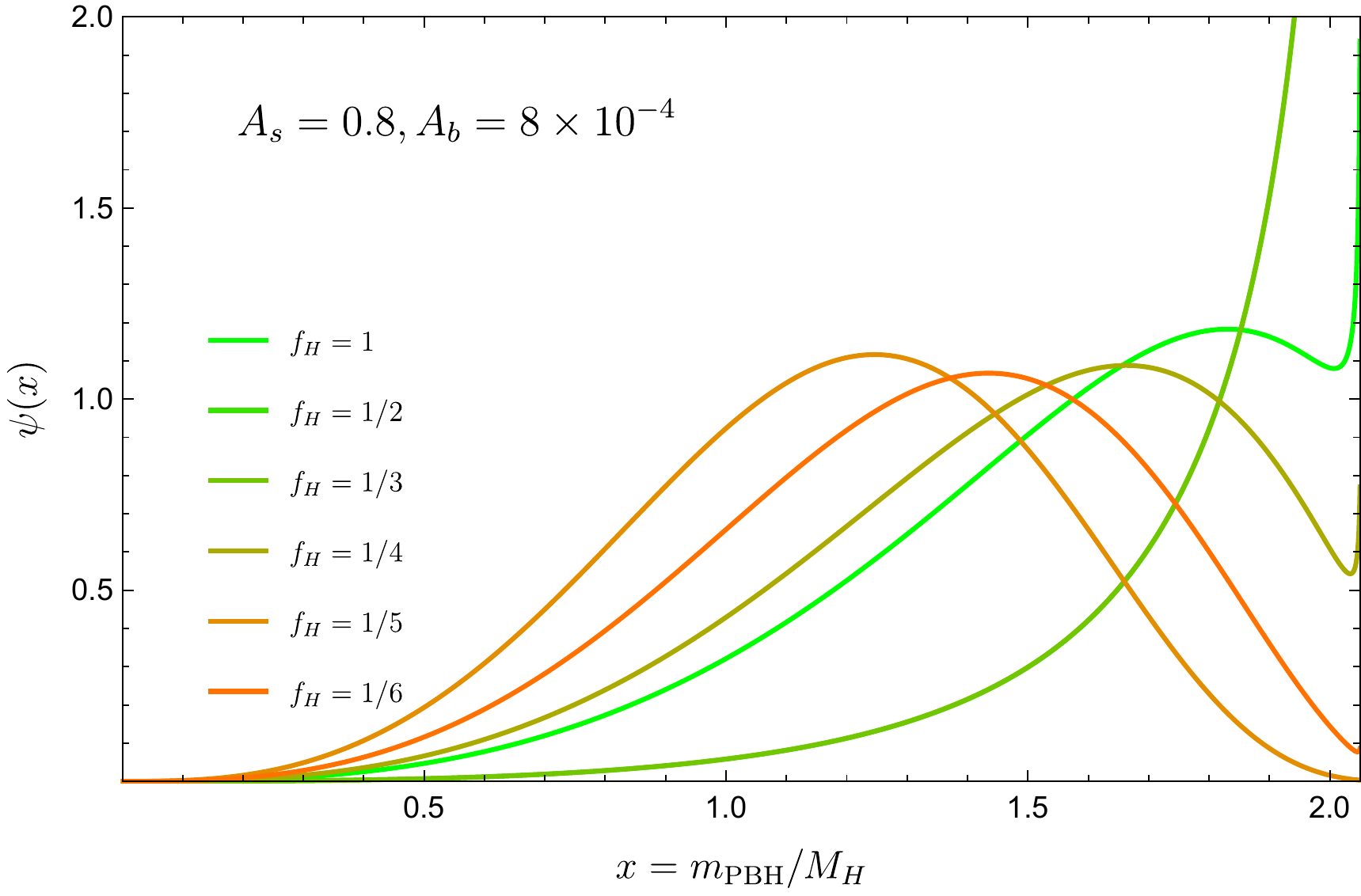}\qquad
  \includegraphics[width=0.45\textwidth]{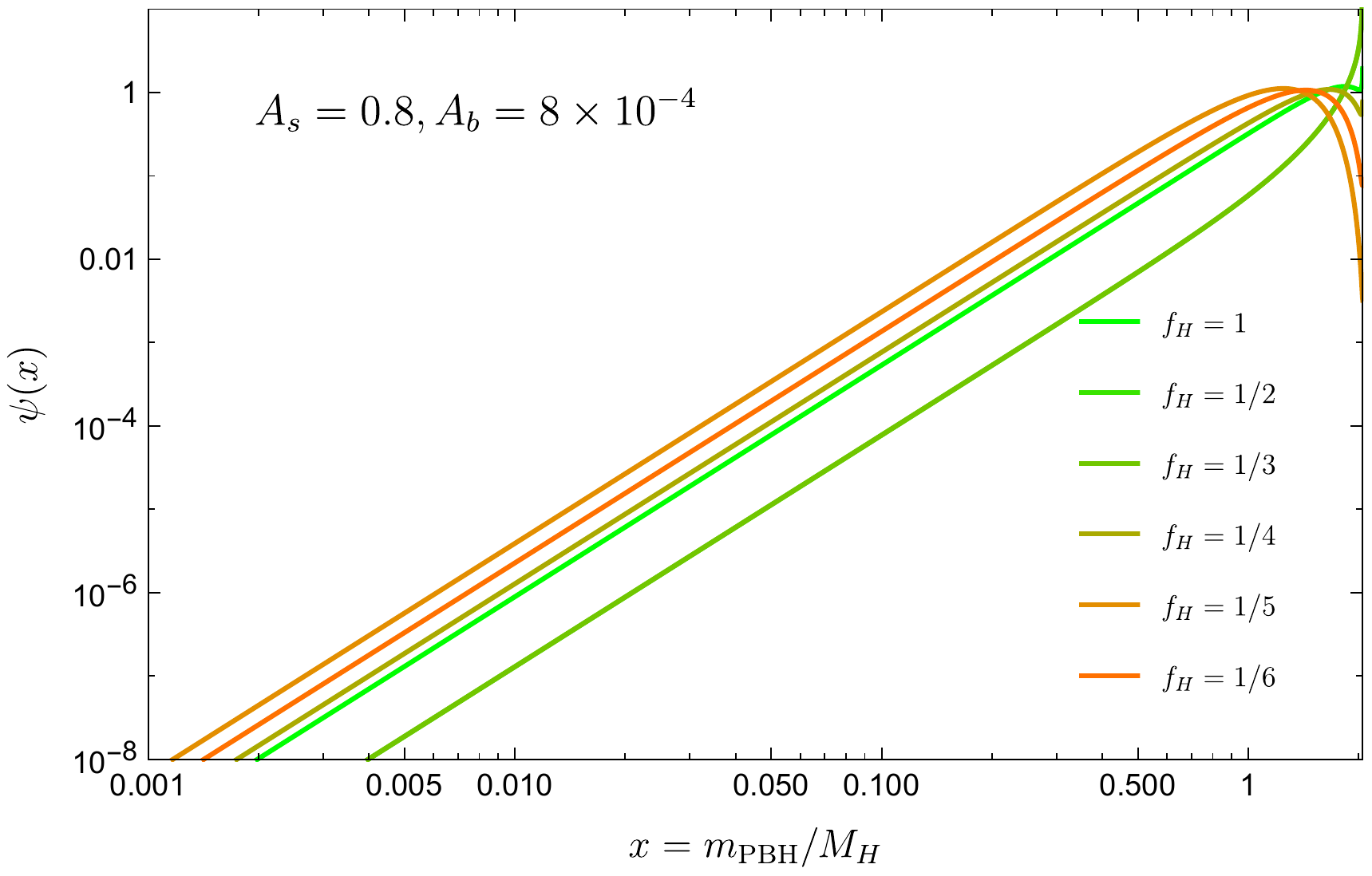}\\
  \includegraphics[width=0.45\textwidth]{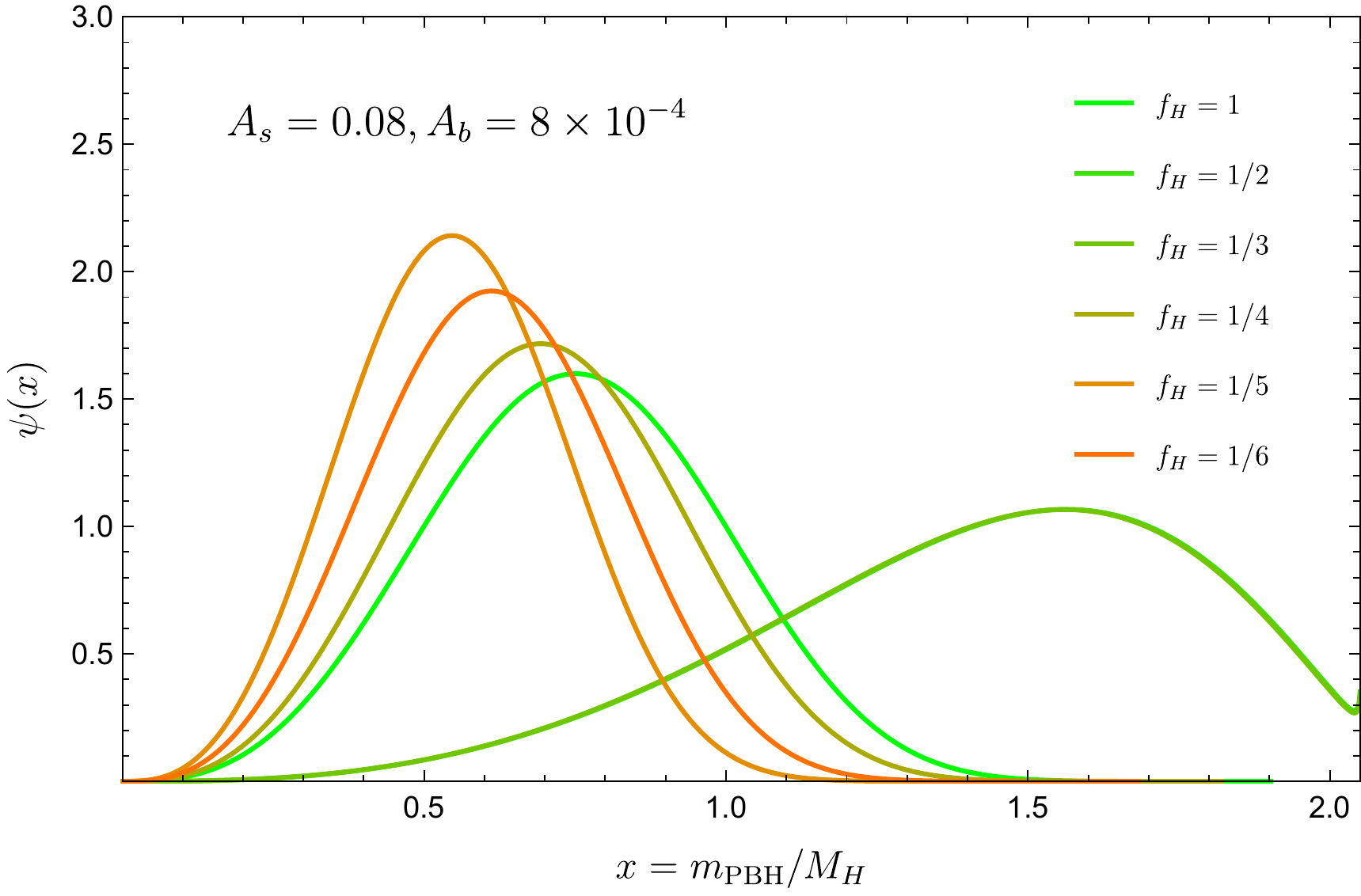}\qquad
  \includegraphics[width=0.45\textwidth]{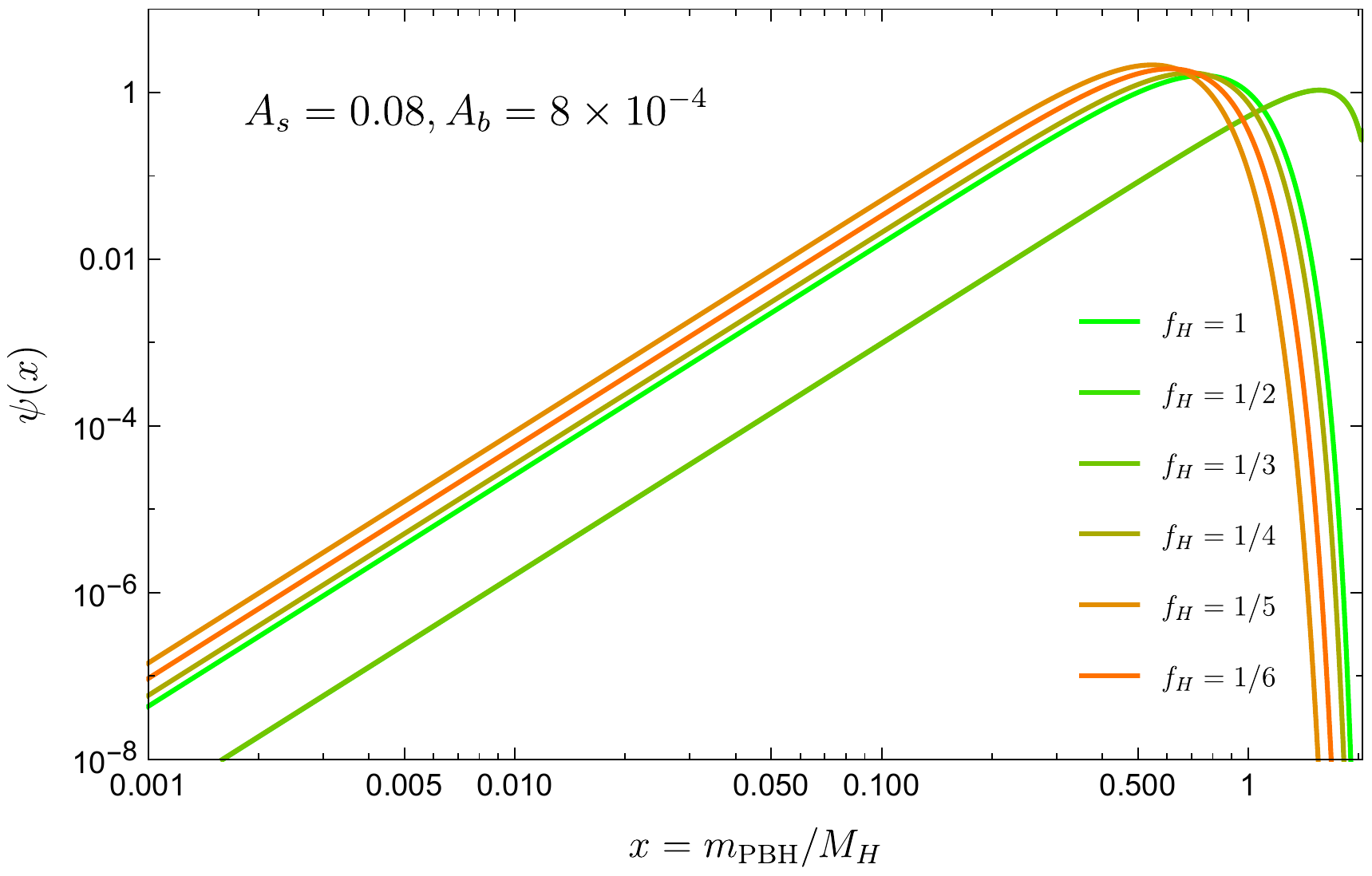}\\
  \includegraphics[width=0.45\textwidth]{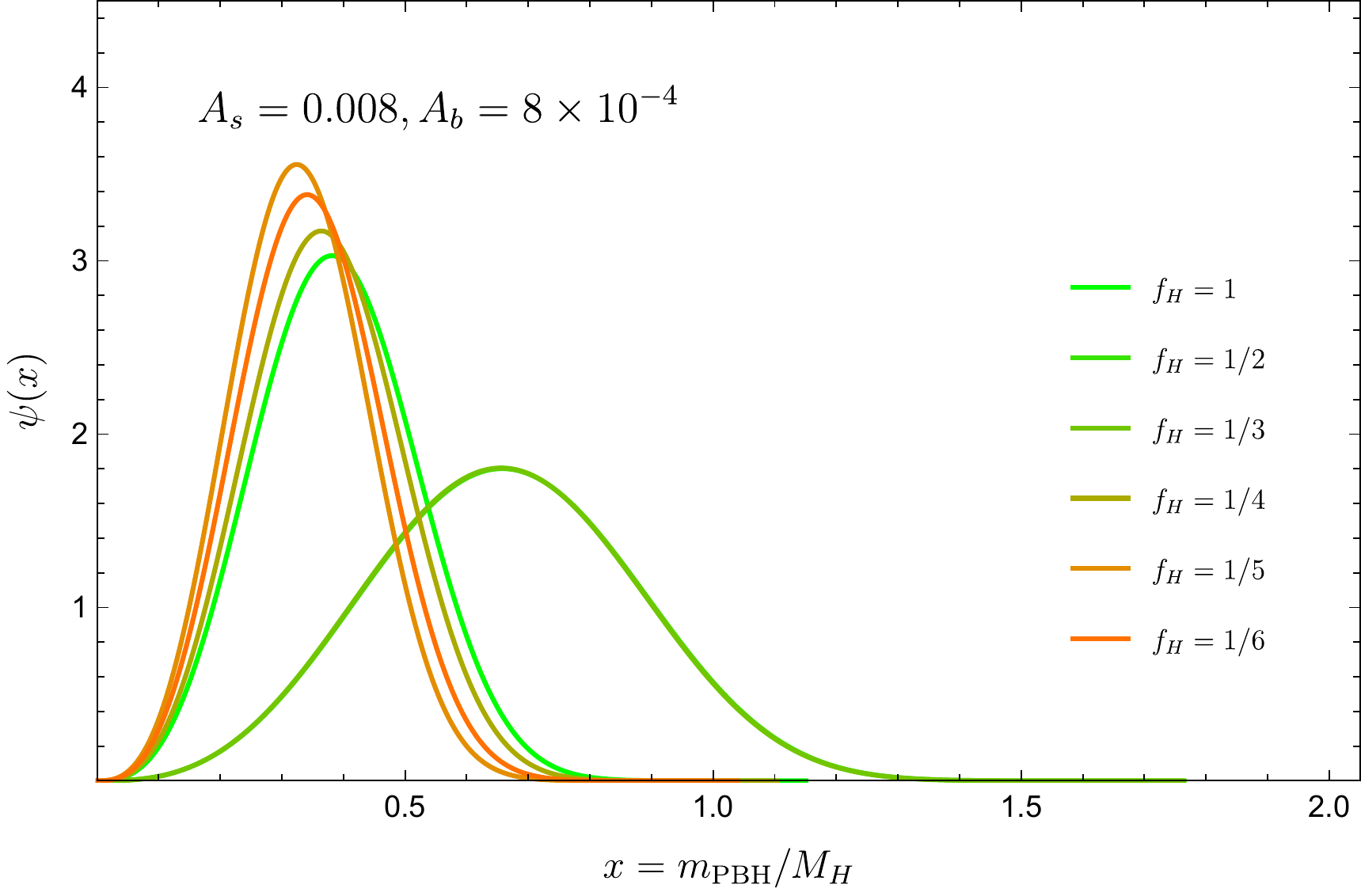}\qquad
  \includegraphics[width=0.45\textwidth]{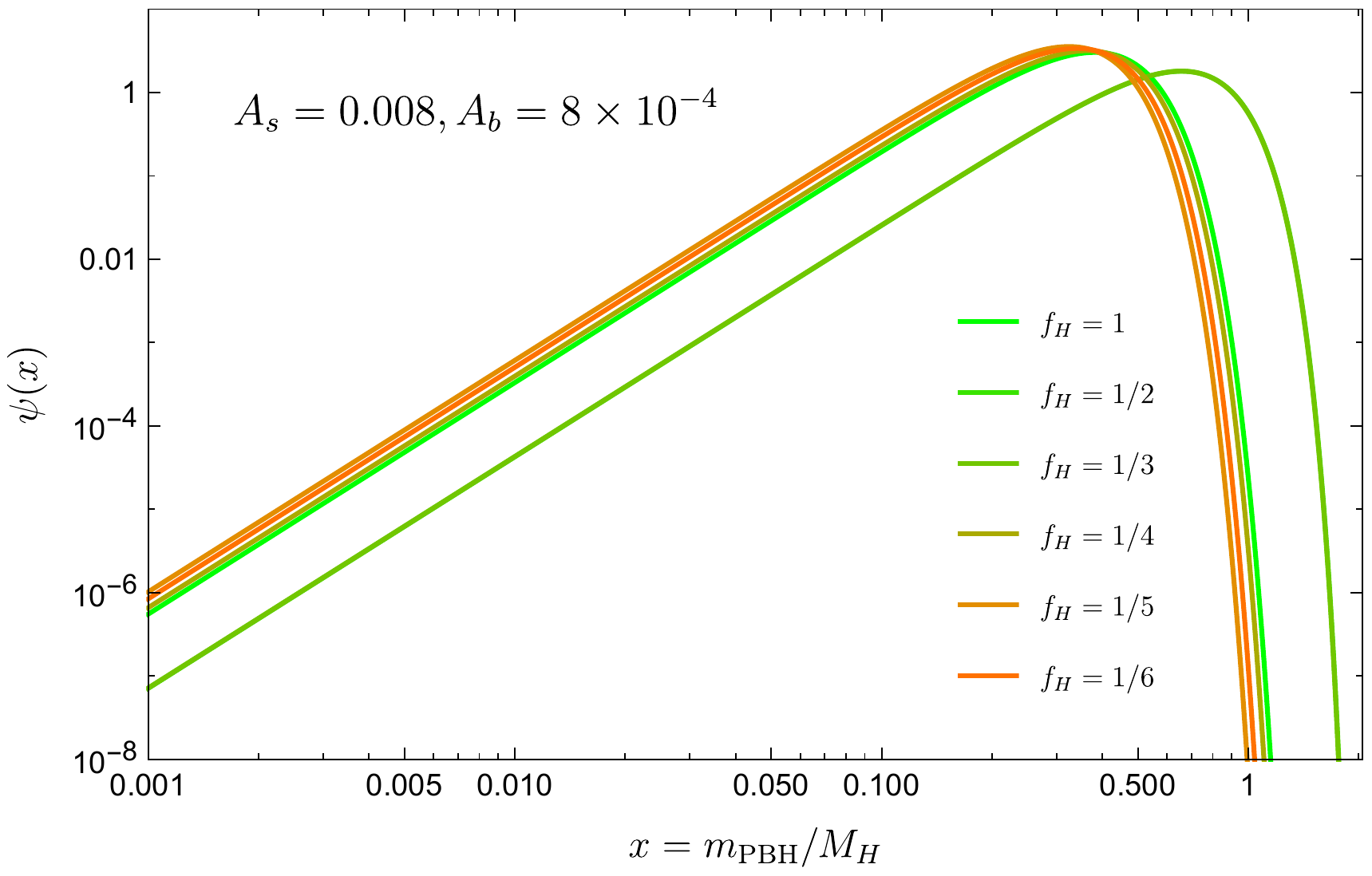}
 \caption{\label{fig:MFOfn} The mass function of PBH for different values of $f_H$, for $A_s=0.8$ (top panels), 0.08 (middle panels) and 0.008 (bottom panels) with $A_b=8\times 10^{-4}$ in all cases. Left column: linear-linear scale. Right column: log-log scale.}
\end{figure}

Figure \ref{fig:MFOfn} shows (on a linear (left) and logarithmic (right) scale)  the mass function of Eq.~(\ref{eq:massfunction}) for different values of the variance $\sigma^2$ induced by $A_s=0.8$, $0.08$, and $0.008$, again at fixed $A_b=8\times10^{-4}$. The log-log plots (Figure \ref{fig:MFOfn} right panels) illustrate how the mass function exhibits a power-law behavior when the normalized PBH mass $x<x_0$, where $x_0$ is the point at which the mass function reaches its local maximum. However, we note that for $A_s=0.8$, $f_H=1/2~\rm{and}~1/3$, the mass function does not have a local maximum, but, rather, it monotonically increases as a function of the PBH mass, while for the other it is exponentially suppressed when $x>x_0$. This behavior can be explained as follows: the exponential function ${\rm exp}\left(-\frac{\delta_1^2(x)}{2\sigma_0^2}\right)$ in Eq.~(\ref{eq:massfunction}) decreases monotonically as $x$ increases for a given $\sigma_0^2$ but with a different decreasing rate which depends sensitively on the value of $\sigma_0^2$. For $f_H^{-1}=2$ or 3 when $A_s=0.8$, the value of $\sigma_0^2\sim0.8$, which is almost an order of magnitude larger than $\sigma_0^2$ at other $f_H^{-1}$, such that the decrease in ${\rm exp}\left(-\frac{\delta_1^2(x)}{2\sigma_0^2}\right)$ is counter-balanced by the increase in the remaining $x$-dependent part in Eq.~(\ref{eq:massfunction}). For other $f_H^{-1}$ values, since the corresponding value of $\sigma_0^2$ is small ($\sim0.1$), the behavior of the mass function at large $x$ is instead mainly determined by the exponential suppression originated from ${\rm exp}\left(-\frac{\delta_1^2(x)}{2\sigma_0^2}\right)$.

\begin{figure}[!t]
\centering  
  \includegraphics[width=0.45\textwidth]{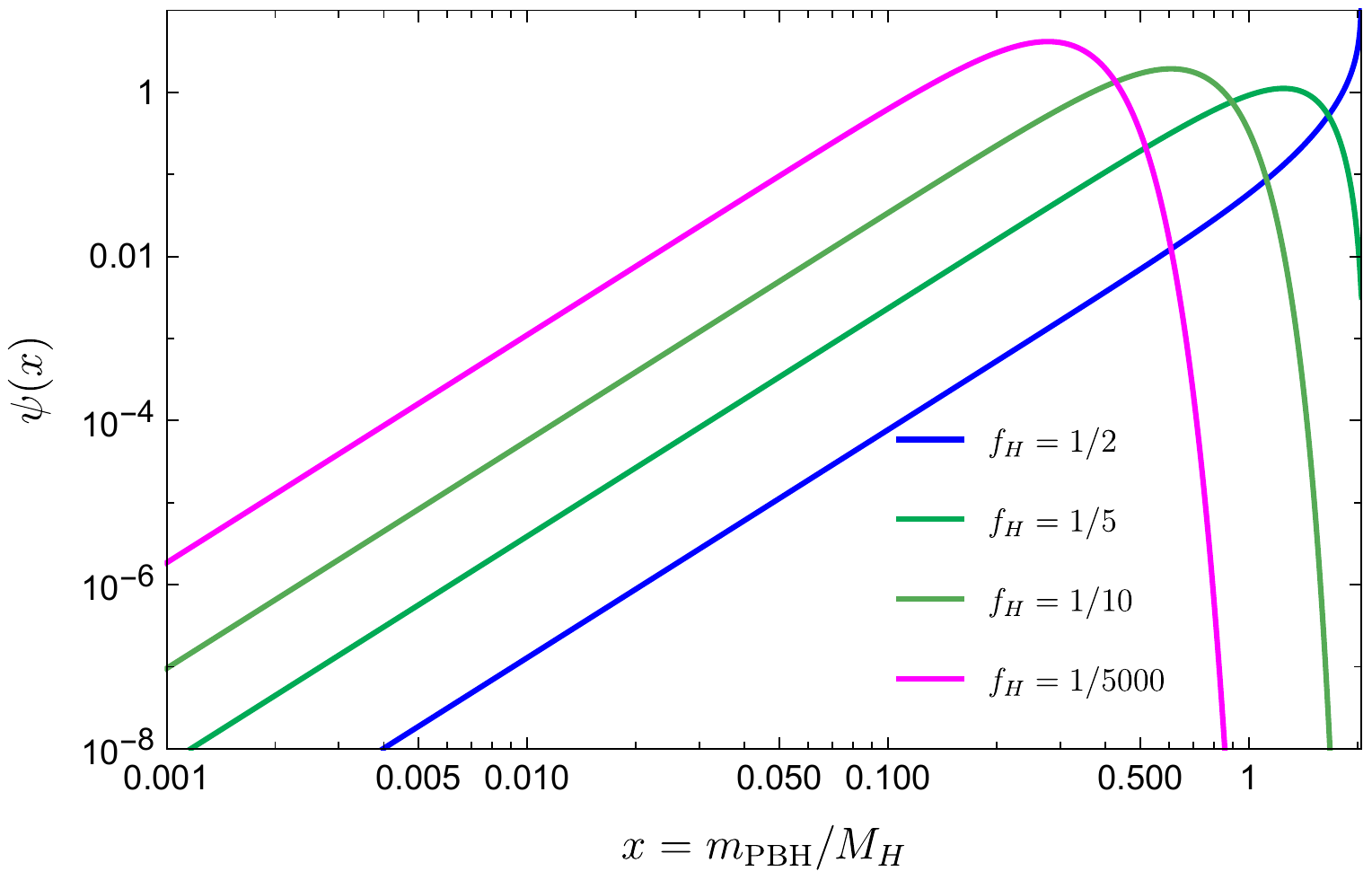}\qquad
  \includegraphics[width=0.45\textwidth]{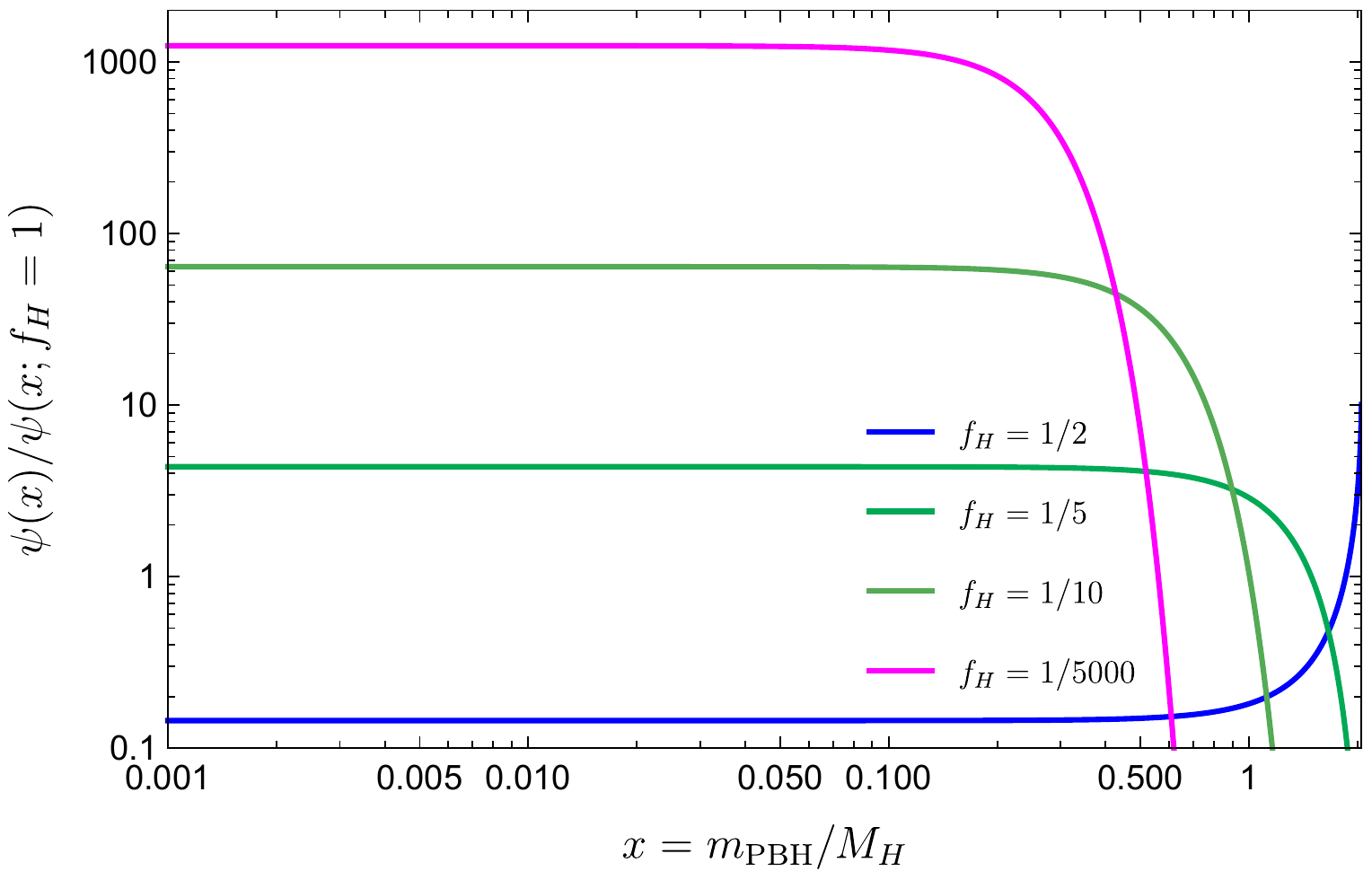}
 \caption{\label{fig:large-n-MF} The mass function (left) and the one normalized by the mass function of $f_H=1$ (right) for $f_H\le 0.1$, $A_s=0.8$, and $A_b=8\times10^{-4}$. }
\end{figure}
We also investigate the mass function at a very small $f_H$, i.e., very large $f_H^{-1}$, ranging from $f_H^{-1}\in[10,5000]$ in Figure \ref{fig:large-n-MF} which shows the mass function at $f_H\ll0.1$ in the left panel and the ratio of mass functions normalized by the ``standard'' mass function corresponding to $f_H=1$ in the right panel. We note that for $f_H\le 1/30$, the mass functions for different values of $f_H$ are self-similar. The reason is that the mass function $\psi(x)$ only depends on the value of $\sigma_0^2=\sigma_0^2(f_H)$ and when $f_H\le 1/30$, $\sigma_0^2\approx0.00085\sim A_b$ does not significantly change.

\section{Anisotropic SGWB energy density power spectrum}\label{sec:aniso}
If two PBHs, separated by a distance  $x$, have an energy density $4M/(3\pi x^3)$ larger than the cosmic radiation energy density $\rho_r$ (since we suppose PBHs are formed deep in the radiation era, radiation is the dominant component of the universe), they decouple from the Hubble flow and become a gravitationally bound binary system. PBH binaries then radiate GWs during the inspiral, merger, and ring-down phases \cite{LIGOScientific:2011hqo}. The incoherent superposition of the unresolved GWs emitted by such PBH binaries contributes to SGWBs across several orders of magnitude in frequency, which advanced LIGO or LISA can detect in the near future \cite{Moore:2014lga}. The SGWB produced by PBH binaries has both an isotropic and an anisotropic component: while only spectral information can be extracted from the former, we focus on the latter, the anisotropic component in seeking a way to ascertain whether the PBH formation history corresponds to a standard horizon-size collapse, or to the aforementioned sub-horizon-collapse scenarios. As such, {\em the scope of the present study is to seek for distinguishable features of PBH binary mergers in the standard versus sub-horizon formation scenarios by studying the anisotropic component}. 

As customary \cite{Allen:1997ad}, we define as the isotropic SGWB energy density spectrum the dimensionless quantity
\be
\bar{\Omega}_{\rm gw}(\nu) \equiv \frac{\nu}{\rho_c}\frac{d\rho_{\rm gw}}{d\nu},
\ee
where $\rho_c$ is the critical energy density of the universe and $\rho_{\rm gw}$ is the SGWB energy density. In a more general situation where the SGWB has directional dependence, the definition is supplemented by
\be
\Omega_{\rm gw}(\nu, {\bf n})\equiv\frac{1}{\rho_c}\frac{d^3\rho_{\rm gw}(\nu,{\bf n})}{d\ln\nu d^2{\bf n}}=\frac{\bar{\Omega}_{\rm gw}}{4\pi}(1+\delta(\nu,{\bf n})),
\ee
where $\rho(\nu,{\bf n})$ is the SGWB energy density at frequency $\nu$ along the line-of-sight direction $\bf n$, $\delta(\nu,{\bf n})$ denotes the anisotropic fluctuations.

\begin{figure}[!t]
\centering  
  \includegraphics[width=0.46\textwidth]{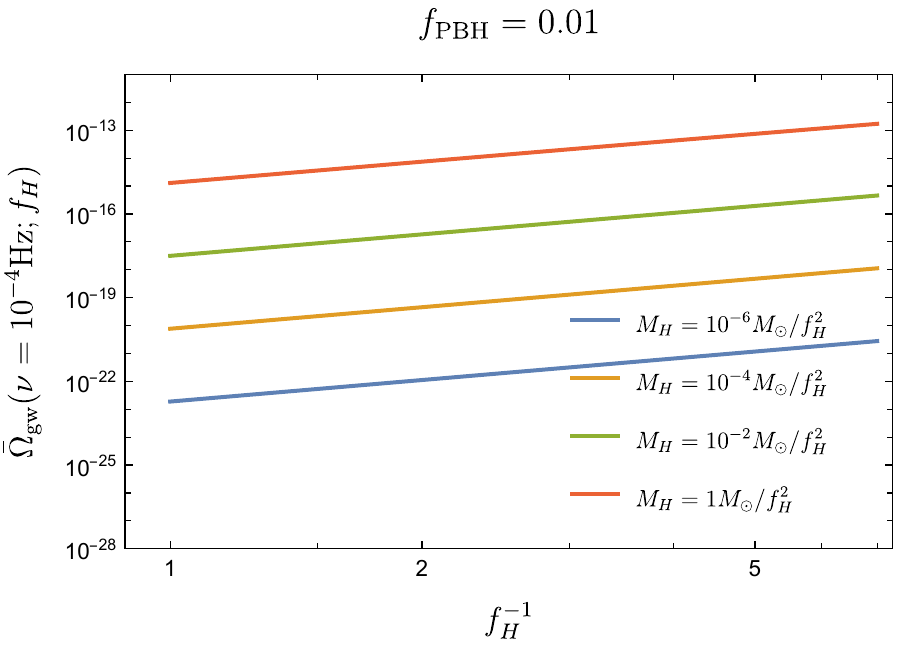}\qquad
  \includegraphics[width=0.46\textwidth]{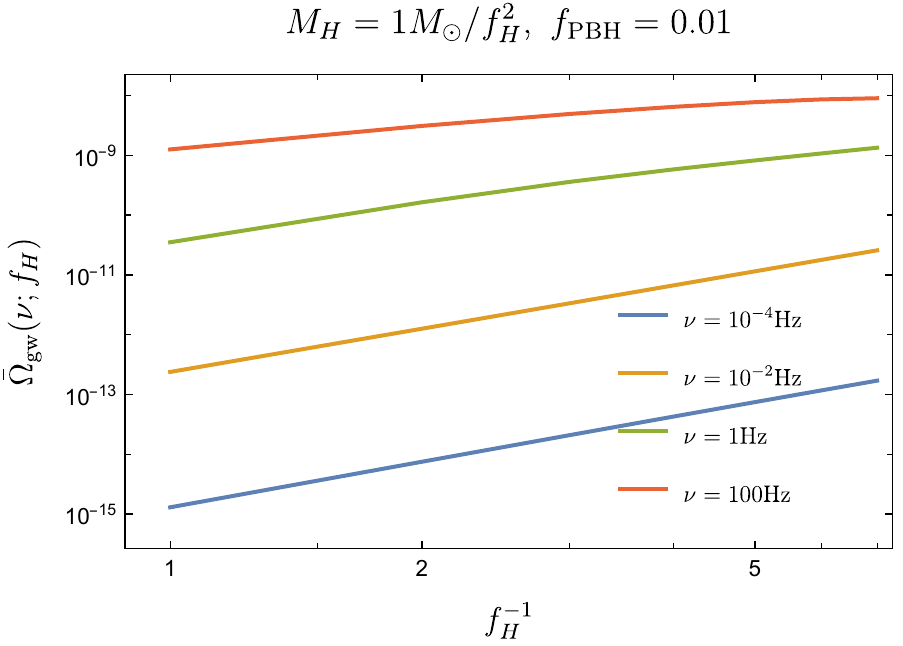}
 \caption{\label{fig:OmegagwOffH} The isotropic component of the SGWB energy spectrum at $\nu=10^{-4}{\rm Hz}$ with different $M_{\rm PBH,max}=M_H=10^{-6}M_\odot/f_H^2$, $10^{-4}M_\odot/f_H^2$, $10^{-2}M_\odot/f_H^2$, and $1M_\odot/f_H^2$ (left), and the stadard case at $M_{\rm PBH,max}=1M_\odot/f_H^2$, for different observed frequencies $\nu=10^{-4}{\rm Hz}$, $10^{-2}{\rm Hz}$, $1{\rm Hz}$, and $100{\rm Hz}$ (right). We fix $f_{\rm PBH}=0.05$.} 
\end{figure}
The isotropic component $\bar{\Omega}_{\rm gw}(\nu)$ produced by PBH binary mergers is given by the following integral:
\be
\bar{\Omega}_{\rm gw}(\nu)=\frac{\nu}{\rho_c}\int_0^{z_*} \frac{dz}{(1+z)H(z)} \int d{\Theta}_s \mathcal{R}(\Theta_s,z)\frac{dE}{d\nu}(\nu_s,\Theta_s),
\ee
where $z$ is redshift, $z_*$ corresponds to the formation time of PBHs, $H(z)=H_0\sqrt{\Omega_r(1+z)^4+\Omega_m(1+z)^3+\Omega_\Lambda}$ is the Hubble parameter, $\mathcal{R}$ is the merger rate of PBH binaries, $dE/d\nu(\nu_s,\Theta)$ is the energy spectrum at source frequency $\nu_s=\nu\times(1+z)$ ,with source parameters $\Theta_s$ describing masses, eccentricity, and so on. The energy spectrum used in our calculation is derived from the inspiral-merger-ringdown waveforms for non-spinning PBH binaries given e.g. in Ref.~\cite{Ajith:2007kx,Ajith:2009bn}. Figure \ref{fig:OmegagwOffH} shows the isotropic SGWB energy density $\bar{\Omega}_{\rm gw}(\nu;f_H)$ as a function of $f_H^{-1}$ at different observed frequencies and different $M_{\rm PBH,max}$, in which we use the merger rate $\mathcal{R}$ (see Eq.~(\ref{eq:mergerrate})) that is to be introduced later to take into account the effect of the sub-horizon formation.

The cross-correlation of the anisotropies in SGWB between two different directions ${\bf n}$ and ${\bf n'}$, $\langle\delta(\nu, {\bf n})\delta(\nu, {\bf n'}) \rangle$, describes the amplitude of the statistical fluctuation at a certain angular scale ${\bf n\cdot n'}$. By decomposing it in a spherical harmonic basis, we have
\be
\langle\delta(\nu, {\bf n})\delta(\nu, {\bf n'}) \rangle = \sum_l \frac{2l+1}{2\pi}C_lP_l({\bf n}\cdot {\bf n'}),
\ee
and the dimensionless angular power spectrum is given by
\be
C_l(\nu)\equiv\frac{2}{\pi}\int d\ln k k^3|\delta_l(k,\nu)|^2,
\ee
where $\delta_l(k,\nu)$ is the Fourier transforms of the spatial anisotropies $\delta(\nu, {\bf n})$, and are given by \cite{PhysRevD.96.103019},
\bea
\label{delta_l_eq}
\delta_l(\nu,k)&=&\frac{\nu}{\bar{\Omega}_{\rm gw}(\nu)\rho_c}\int_{\eta_*}^{\eta_0}d\eta \mathcal{L}(\eta,\nu_s)\nonumber\\
&~&\times\Bigg\{ \left[4\Psi_k(\eta)+4\Pi_k(\eta)+b\delta_{m,k}(\eta)\right]j_l(k\Delta\eta)-2kv_k(\eta)j_l'(k\Delta\eta)-6\int_{\eta_0}^\eta d\eta'\dot{\Psi}_k(\eta')j_l(k\Delta\eta')\nonumber\\
&~&+\frac{\partial \ln(dE/d\nu_s)}{\partial\ln\nu_s}\left[-\Psi_k(\eta)j_l(k\Delta\eta)-\Pi_k(\eta)j_l(k\Delta\eta)+kv_k(\eta)j_l'(k\Delta\eta)+2\int_{\eta_0}^\eta d\eta'{\Psi}_k(\eta')j_l(k\Delta\eta')\right]\Bigg\},
\eea
where $\mathcal{L}(\eta,\nu_s)=a(\eta)\int d\Theta_s \mathcal{R}(\Theta_s,z(\eta))\frac{dE}{d\nu}(\nu_s,\Theta_s)$ is the GW luminosity of the PBH binary merger, the galaxy bias $b=1$ for PBH, $\Psi_k\approx\phi_k$ and $\Pi_k=0$ are two scalar gravitational potentials, $\delta_{m,k}$ is the matter density contrast, $v_k$ is the velocity of matter, the dot represents the derivative with respect to conformal time $\eta$. The corresponding transfer functions are calculated numerically by solving the linearized cosmological perturbation equations in CLASS, and the primordial scalar fluctuation power spectrum is $P_s(k)=\mathcal{A}_s^2k^{-3}(k/k_0)^{n_s-1}$, where $\mathcal{A}_s=2.215\times10^{-9}$, $n_s=0.96$, and $k_0=0.05{\rm Mpc}^{-1}$ are chosen from the Planck 2018 results \cite{Planck:2018vyg}, and $j_l$ is the spherical Bessel function. 

By comparing the numerical results of $\delta_l$ in Eq.~(\ref{delta_l_eq}) with or without including the $v_k$'s and $\phi_k$'s terms, we find that only the term with the matter density contrast $\delta_m$ is dominant in the calculation, so it can be approximated by
\be
\label{delta_l}
\delta_l(\nu,k)=\frac{\nu}{\bar{\Omega}_{\rm gw}(\nu)\rho_c}\int_{\eta_*}^{\eta_0}d\eta \mathcal{L}(\eta,\nu_s)\delta_{m,k}(\eta)j_l(k\Delta\eta).
\ee

We assume that the merger rate of PBH is given by the functional form \cite{Raidal:2018bbj},
\be
\label{eq:mergerrate}
\mathcal{R}(\Theta_s,\tau)=1.6\times 10^6\text{Gpc}^{-3}\text{yr}^{-1}~ f_{\rm PBH}^{53/37}\eta(m_1,m_2)^{-34/37}\left(\frac{m_1+m_2}{M_\odot}\right)^{-32/37}\left(\frac{\tau}{t_0}\right)^{-34/37}\Psi(m_1)\Psi(m_2)S(f_{\rm PBH}),
\ee
where $f_{\rm PBH}$ (not to be confused with $f_H$) is the fraction of PBH at the present time $t_0$, $\eta(m_1,m_2)=m_1m_2/(m_1+m_2)^2$ is the symmetric factor for the binary, $\Psi=\Psi(m;f_H)$ is the mass function which is defined in the previous section, and $S(f_{\rm PBH})=\left(\frac{5f_{\rm PBH}^2}{6\sigma_M^2}\right)^{21/74}U\left(\frac{21}{74},\frac{1}{2},\frac{5f_{\rm PBH}^2}{6\sigma_M^2}\right)$ is the suppression factor with $U$ the confluent hypergeometric function. Notice that we assume that the form of the merger rate is unchanged in the sub-horizon formation scenario, i.e., $\Psi(m;f_H)$ can be evaluated when $f_H<1$, which is strongly physically motivated.

\begin{figure}
\centering  
  \includegraphics[width=0.52\textwidth]{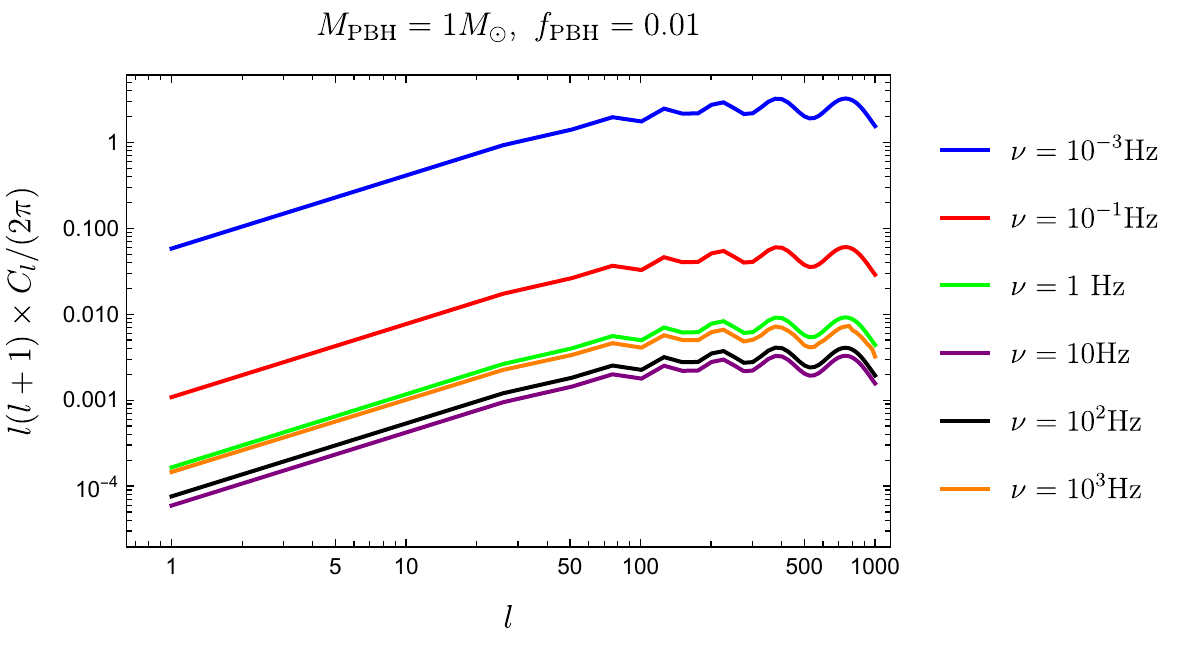}\quad
  \includegraphics[width=0.4\textwidth]{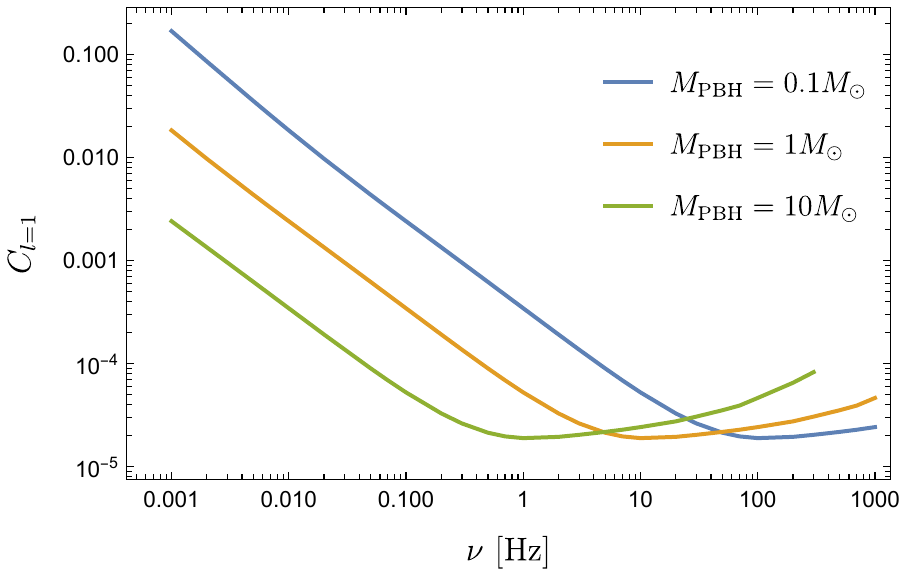}
 \caption{\label{fig:cl-multinu} The angular power spectrum at multiple frequencies at a fixed PBH mass $M_{\rm PBH}=1M_\odot$ (left). The relative amplitude of angular power spectrum at $l=1$ as a function of frequency $\nu$ for three different PBH masses $M_{\rm PBH}=0.1,1,10~M_\odot$ (right).} 
\end{figure}
Figure \ref{fig:cl-multinu}, left, shows the typical angular power spectrum of the SGWB anisotropic component for PBH binary mergers in the standard scenario. We choose a Dirac-$\delta$ mass function in the merger rate such that $M_{\rm PBH}=M_H=1M_\odot$, and thus PBHs are formed deep in the radiation-dominated era with $z_{\rm form}=1.62\times10^{12}$.
The right panel shows the frequency dependence of the $l=1$ dipole as a function of frequency for various choices of the PBH mass $M_{\rm PBH}=0.1,1,10~M_\odot$, illustrating how larger chirp masses, $M_c=2^{-1/5}M_{\rm PBH}$ for a binary PBH with the same mass, induce a lower dipole amplitude at low frequency and a larger amplitude at higher frequencies, as expected.

\begin{figure}[!t]
\centering  
  \includegraphics[width=0.45\textwidth]{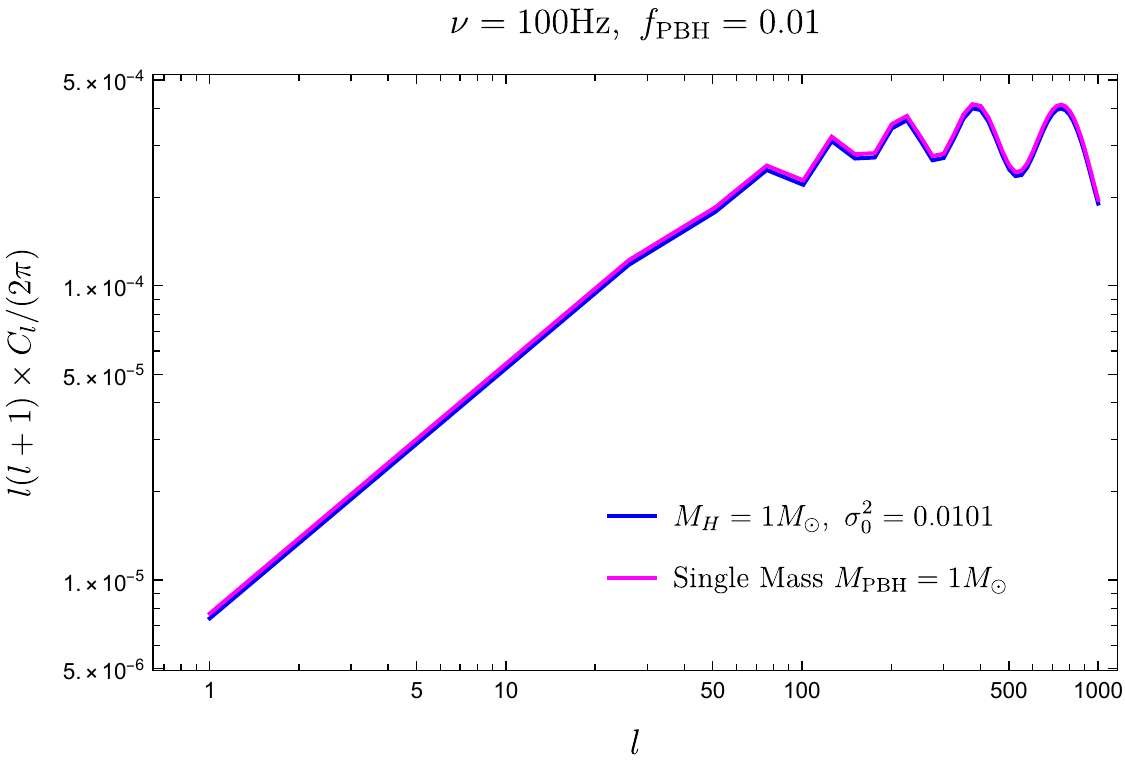}\qquad
  \includegraphics[width=0.43\textwidth]{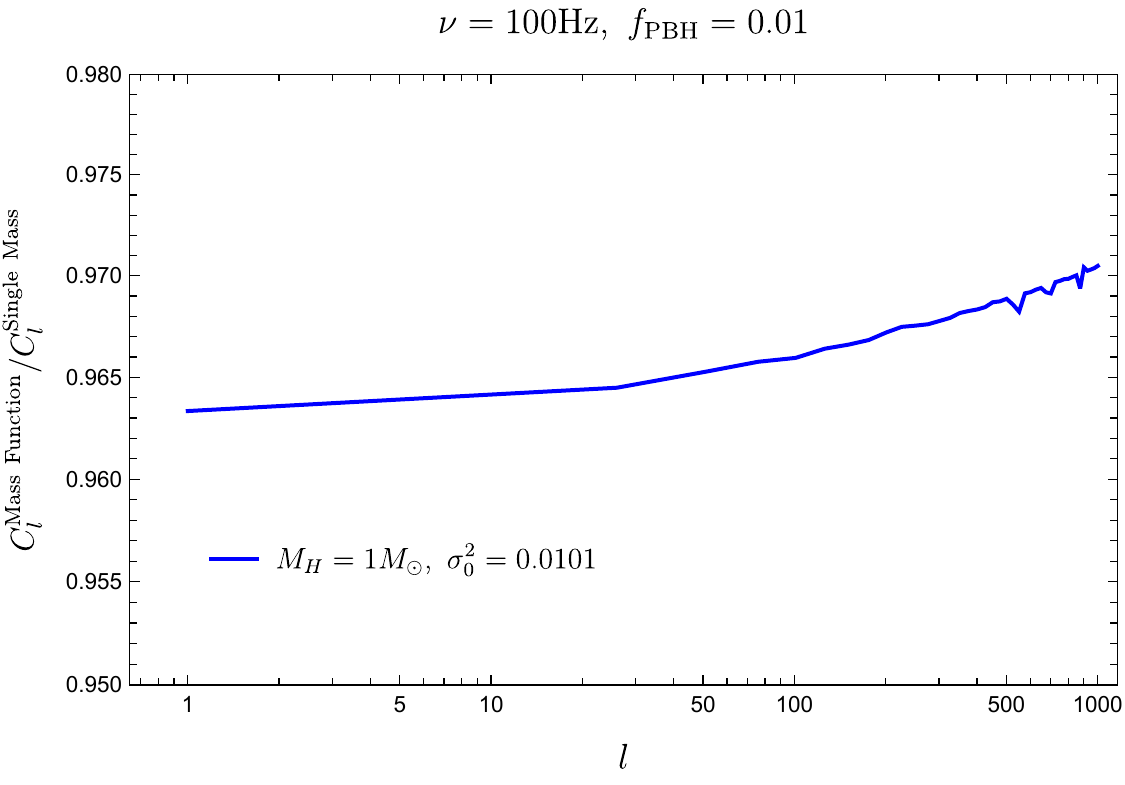}
 \caption{\label{fig:cl-100hz-comp} The angular power spectrum $C_l$ at 100 Hz with PBH masses given by a mass function ($f_H=1$ and $\sigma_0^2=0.0101$ such that $f_{\rm PBH}=0.01$) and by a fixed PBH mass $M_{\rm PBH}=1M_\odot$ (left). The ratio of $C_l$ between the mass function cases and the single mass case (right). 
 }
\end{figure}

In Figure \ref{fig:cl-100hz-comp} we compare the angular power spectrum obtained before to that corresponding to a mass function arising from a value of $\sigma_0^2$ such that $f_{\rm PBH}=0.01$ at a frequency $\nu=100{\rm Hz}$ (the right plot shows the ratio of the curves at a given $\sigma_0^2$ to the standard case). There are no large deviations between the angular power spectra besides an overall amplitude shift due to the different weighting on the PBH mass function. The single mass case has a Dirac-$\delta$ peak mass function, while the other has a mass function determined by $\sigma_0^2$.

\begin{figure}[!t]
\centering  
  \includegraphics[width=0.8\textwidth]{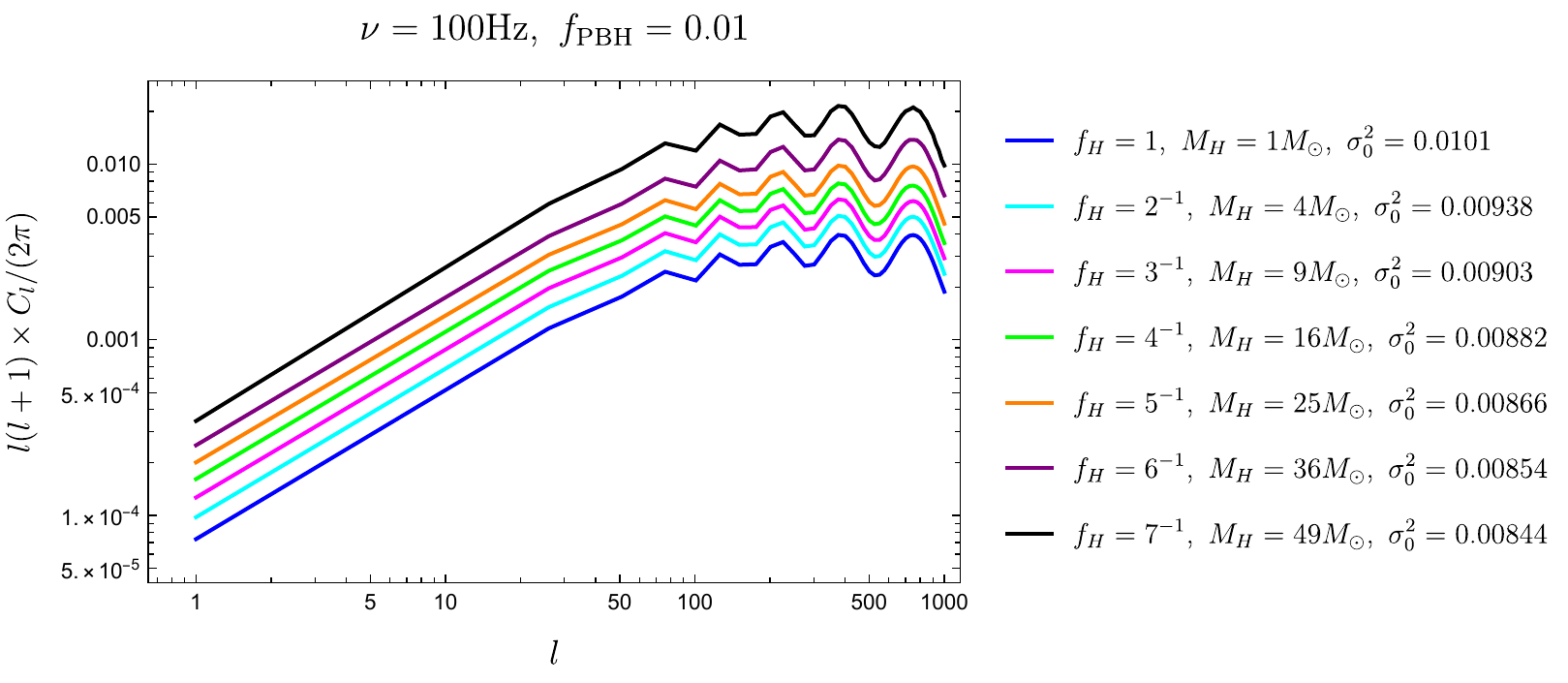}
 \caption{\label{fig:cl-100hz} The angular power spectrum at 100 Hz for different $f_H$.} 
\end{figure}

\begin{table}[h!]
\begin{center}
\begin{tabular}{||c | c | c | c | c | c | c | c | c | c | c | c ||} 
 \hline
 $f_H^{-1}$ & 1 & 1.2 & 1.4 & 1.6 & 1.8 &  2 & 3 & 4 & 5 & 6 & 7  \\ [0.5ex] 
 \hline
 $\bar{\Omega}_{\rm gw}/ 10^{-9}$ & 1.26 & 1.61 & 1.98 & 2.36 & 2.73 & 3.11 & 4.93 & 6.49 & 7.77 & 8.63 & 9.04\\
 \hline
 $\sigma_0^2$ & 0.0101 & 0.00988 & 0.00972 & 0.00959 & 0.00948 &  0.00938 & 0.00903 & 0.00882 & 0.00866 & 0.00854 & 0.00844 \\
 \hline
 $\sigma_1^2$ & 0.0101 & 0.0142 & 0.0191 & 0.0245 & 0.0307 &  0.0375 & 0.0813 & 0.141 & 0.217 & 0.307 & 0.413\\ 
 \hline
 $A_s$ & 0.0668 & 0.0357 & 0.0219 & 0.0150 & 0.0112 &  0.00913 & 0.00885 & 0.0736 & 0.149 & 0.103 & 1.76\\ [1ex]
 \hline
\end{tabular}
\end{center}
 \caption{Values of the isotropic SGWB energy density $\bar{\Omega}_{\rm gw}$ at 100 Hz, the zeroth order and first order width, and the amplitude of Dirac-$\delta$ curvature power spectrum for a fixed $f_{\rm PBH}$ and different $f_H^{-1}$.  }
\label{table:1}
\end{table}
To study the effects of $f_H\neq1$, we study the SGWB from an $f_H$-dependent mass function $\Psi(m;f_H)$ to investigate the effect of sub-horizon formation on the angular power spectrum.  Figure \ref{fig:cl-100hz} shows the angular power spectrum at 100 Hz for $f_H=m^{-1}$, $1\le m\le7$ and $m\in\mathbb{N}$. 
The amplitude of the angular power spectrum increases as $f_H^{-1}$ increases when $f_H^{-1}\in[1,6]$, but slightly decreases when $f_H^{-1}\in(6,7]$. Table \ref{table:1} summarizes the isotropic SGWB energy density values, $\sigma_0^2$, $\sigma_1^2$, and $A_s$ required to match a given $f_{\rm PBH}=0.01$ for different $f_H$ (including some fractional values of $m$). 


\begin{figure}[!t]
\centering  
  \includegraphics[width=0.45\textwidth]{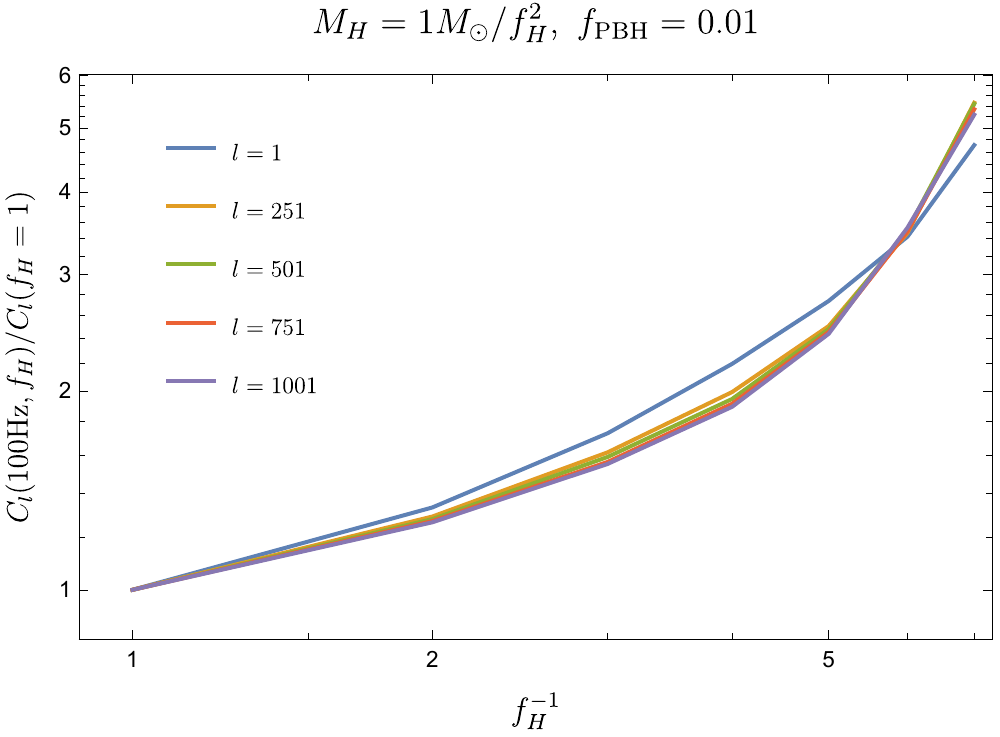}\qquad
  \includegraphics[width=0.45\textwidth]{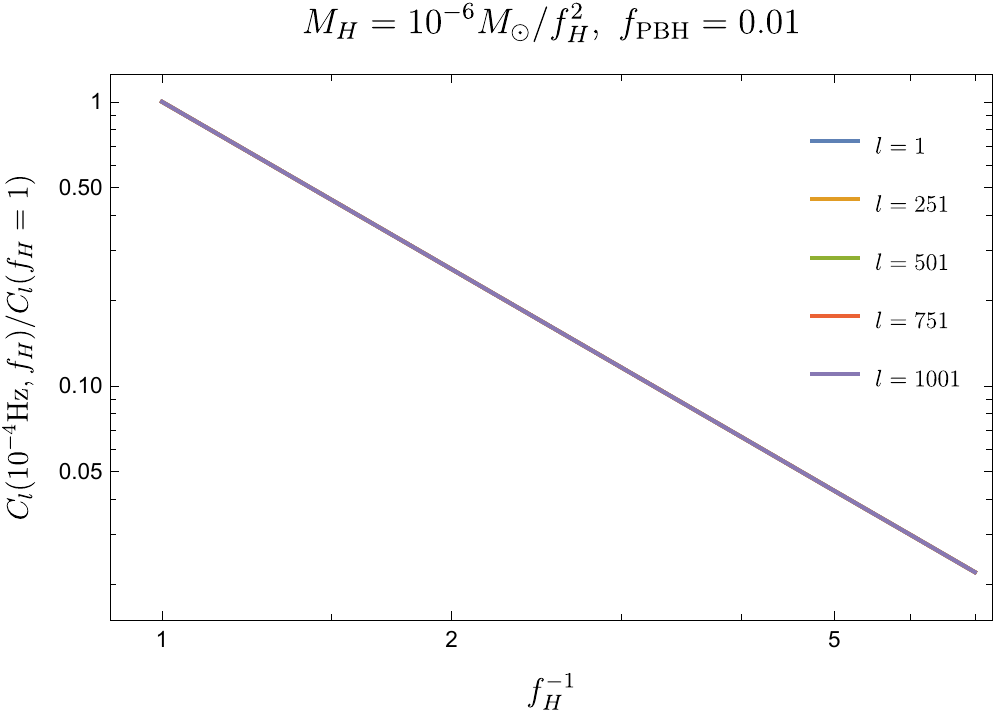}
 \caption{\label{fig:Cl-100hz-frac-n} The angular power spectra of $l=1,~251,~501,~751,~{\rm and}~1001$ as a function of $f_H^{-1}$, normalized by $C_{l}(f_H=1)$ for the corresponding $l$, for \{1$M_\odot$,100Hz\} (left) and \{$10^{-6}M_\odot$,$10^{-4}$Hz\} (right).} 
\end{figure}

We are also interested in the effect of sub-horizon formation on the angular power spectrum $C_l$ at different multipole $l'$s. We show the normalized $C_l$ (i.e. the ratio $C_l(f_H)/C_l(f_H=1)$) in Figure \ref{fig:Cl-100hz-frac-n}, for different parameter sets $\{M_{H}f_H^2,\nu\}$, in which the first parameter sets the horizon mass when PBHs form for different $f_H$ and the second parameter is the observed frequency of the angular power spectrum. 
The left panel employs \{1$M_\odot$,100Hz\} while the right panel \{$10^{-6}M_\odot$,$10^{-4}$Hz\}. We also sampled \{$10^{-2}M_\odot$,1Hz\} and \{$10^{-4}M_\odot$,$10^{-2}$Hz\}, and found almost identical results as for  \{$10^{-6}M_\odot$,$10^{-4}$Hz\}.

The choice of the parameter sets makes the observed frequency below the merger frequency of the IMR waveform at a redshift $z=0$. We do not find a significant difference for $C_l$ at different multipole $l$ when choosing different $f_H$. The behavior of the $C_l$ ratio at 100Hz is different from the other cases we studied ($\nu=10^{-2}$Hz, $10^{-4}$Hz, and $10^{-6}$Hz), which is, in turn, similar to the behavior of $C_l$ shown in Figure \ref{fig:cl-multinu}. There exists a turn-around frequency $\nu_{\rm ta}$ for a given $f_H$ and $M_{\rm PBH}$ or effectively $M_c$ such that $C_l$ decreases as $\nu$ increases when $\nu<\nu_{\rm ta}$, while $C_l$ increases when $\nu\ge\nu_{\rm ta}$. Similarly, for a fixed observed frequency $\nu$ and multiple $f_H$ and $M_{H}$ or equivalently $M_{\rm PBH,max}=2.05M_H$, we can define a minimal turn-around frequency $\nu_{\rm ta, min}=\min\{\nu_{\rm ta}(M_{H,i}f_{H,i}^2)\}$ and a maximal turn-around frequency $\nu_{\rm ta, max}=\max\{\nu_{\rm ta}(M_{H,i}f_{H,i}^2)\}$. If $\nu<\nu_{\rm ta, min}$, $C_l$ will generally decrease as $M_{H}$ increases. If $\nu>\nu_{\rm ta, max}$, $C_l$ will generally increase as $M_{H}$ increases.

\begin{figure}[!t]
\centering  
  \includegraphics[width=0.46\textwidth]{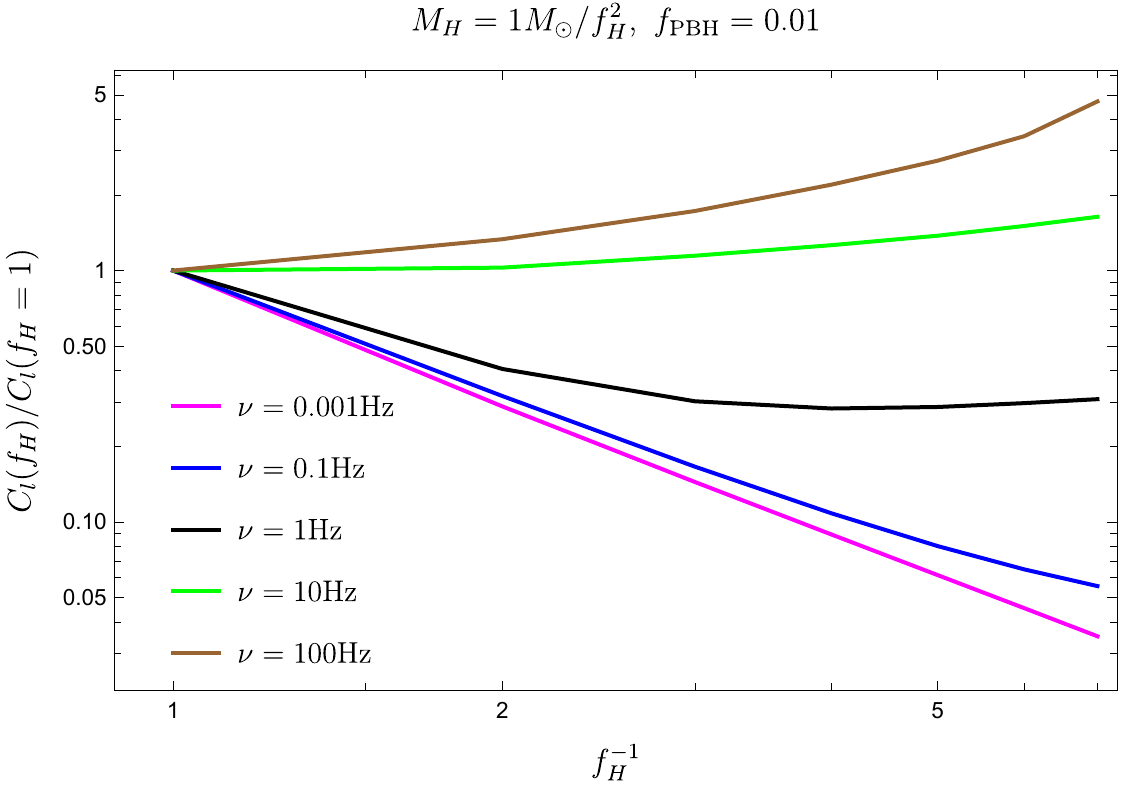}\qquad
  \includegraphics[width=0.46\textwidth]{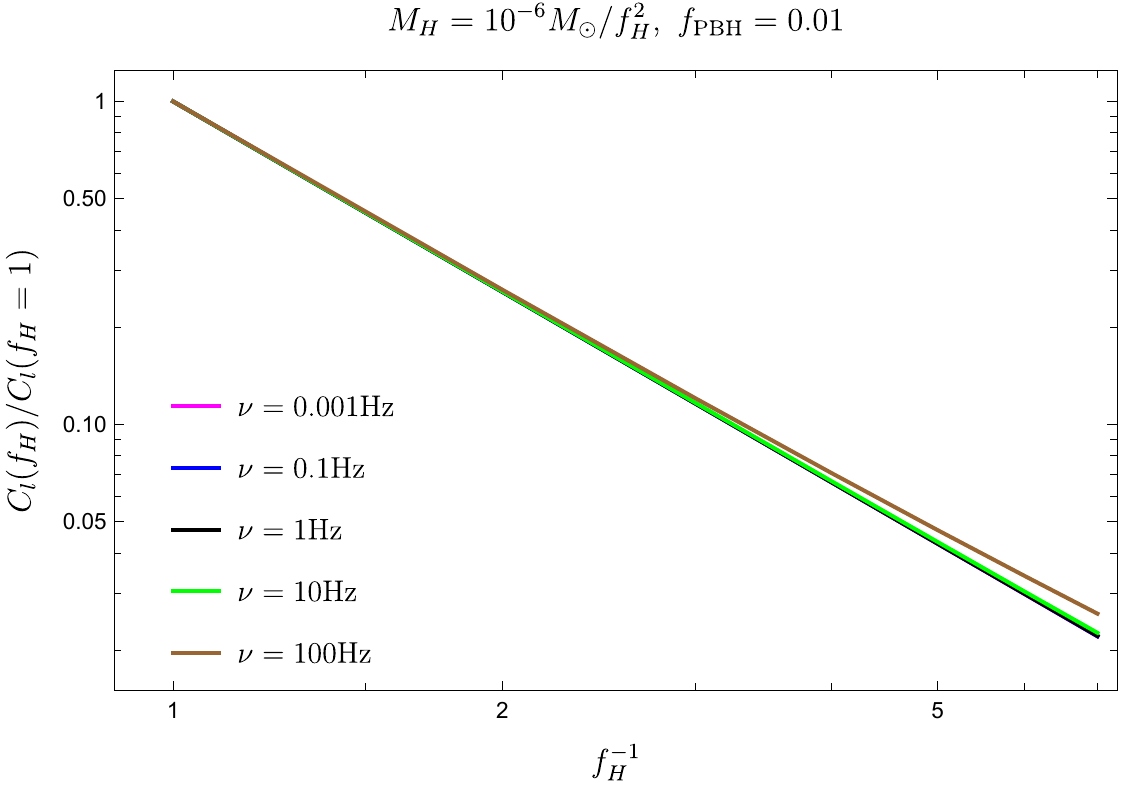}
 \caption{\label{fig:Cl-ratio-frac-n-diff-nu} The angular power spectra for $l=1$ as a function of $f_H^{-1}$ normalized by $C_l(\nu;f_H=1)$ for different frequencies. The left panel is using $M_{H}=1M_{\odot}/f_H^2$, while the right panel is using $M_{H}=10^{-6}M_{\odot}/f_H^2$.}
\end{figure}
The ratio $C_l(f_H)/C_l(f_H=1)$ is an $f_H$-dependent and frequency-dependent object, as shown in Figure \ref{fig:Cl-ratio-frac-n-diff-nu}. 
To analytically understand this behavior, we first rewrite $j_l(x)$ using the Limber approximation,
\be
\label{limber-approx}
j_l(x)\approx\sqrt{\frac{\pi}{2l+1}}\delta(x-(l+0.5)),
\ee
where $\delta(x)$ is the Dirac $\delta$ function. Applying the Limber approximation to  Eq. \ref{delta_l} by replacing $x=k\Delta\eta=k(\eta-\eta_0)$, we can integrate the conformal time integral by using the Dirac $\delta$-function, which gives
\be
\delta_l(\nu,k)=\sqrt{\frac{\pi}{2l+1}}\frac{\nu}{\bar{\Omega}_{\rm gw}(\nu)\rho_c}\frac{\mathcal{L}(\eta_*,\nu_s)}{1+z_*} \frac{\delta_{m,k}(\eta_*)}{k},
\ee
where $\eta_*=\eta_0-\frac{l+0.5}{k}$ is the conformal time determined by different sets of $\{l,k\}$ such that the argument of the Dirac $\delta$-function in Eq.~(\ref{limber-approx}) is zero, and $z_*$ is the redshift evaluated at $\eta_*$. The angular power spectrum is thus simplified in the following,
\be
C_l(\nu)=\frac{2}{2l+1}\left(\frac{\nu}{\bar{\Omega}_{\rm gw}(\nu)\rho_c}\right)^2\int dk\frac{\mathcal{L}^2(\eta_*,\nu_s)}{(1+z_*)^2} \delta_{m,k}^2(\eta_*).
\ee

For a fixed $l$ and a fixed observed frequency $\nu$ but different $f_H$, the ratio of $C_l$ is given by,
\be
\label{eq:clratio}
\frac{C_l(\nu;f_H)}{C_l(\nu;f_H=1)}=\left(\frac{\bar{\Omega}_{\rm gw}(\nu;f_H=1)}{\bar{\Omega}_{\rm gw}(\nu;f_H)}\right)^2\left(\frac{I(\nu;f_H)}{I(\nu;f_H=1)}\right)^2,
\ee
where $I(\nu;f_H)=\int dk\frac{\mathcal{L}^2(\eta_*,\nu_s;f_H)}{(1+z_*)^2} \delta_{m,k}^2(\eta_*)$. 

In the present scenario under consideration, the horizon mass is increased by a factor of $f_H^{-2}$ when PBHs are formed in the radiation-dominated era, and thus the maximal PBH mass is also increased by the same factor. Note that in early matter domination, the horizon mass would also be increased, but by the smaller factor $f_H^{-3/2}$; in a general cosmology where the equation of state of the dominant species is $P=w\rho$, the horizon mass varies as $f_H^{-3(w+1)/2}$. The mass function also depends on $f_H$ since the width of the distribution PDF$(\delta_1)$ depends on $f_H$, i.e., $\sigma_0=\sigma_0(f_H)$ if $A_s$ is given. Since PDF$(\delta_1)$ is a Gaussian distribution, the shape of the distribution is sensitive to the change in $\sigma_0$  Therefore, the merger rate and the GW energy spectrum depend on $f_H$ non-trivially in general: For example, the merger rate depends on the total mass of the binary system by $M_{\rm tot}^{-32/37}$, in which we suppose the two BHs have the same mass, i.e., $\eta(m_1,m_2)=0.25=\eta_{\rm max}$, but for different $f_H$ the mass function is different, and it can peak at different $x=M_{\rm PBH}/M_H$ (see Figure \ref{fig:MFOfn}), so only using $f_H$ cannot analytically predict the behavior of the merger rate. {\it However, by requiring a fixed $f_{\rm PBH}$, the $\sigma_0^2$ is almost determined, and varying $f_H$ only affects the horizon mass when PBHs form and PBH formation time.} For the GW energy spectrum, varying $f_H$ will directly vary the spectrum amplitude -- the spectrum amplitude is proportional to $M_{\rm tot}^{5/3}$ if $m_1=m_2$ for the inspiral phase where $\nu$ is smaller than the merger frequency, and the shape of the spectrum, e.g., the cutoff frequency shifts to a lower frequency for a larger chirp mass $M_c$.

\begin{figure}[!t]
\centering  
  \includegraphics[width=0.45\textwidth]{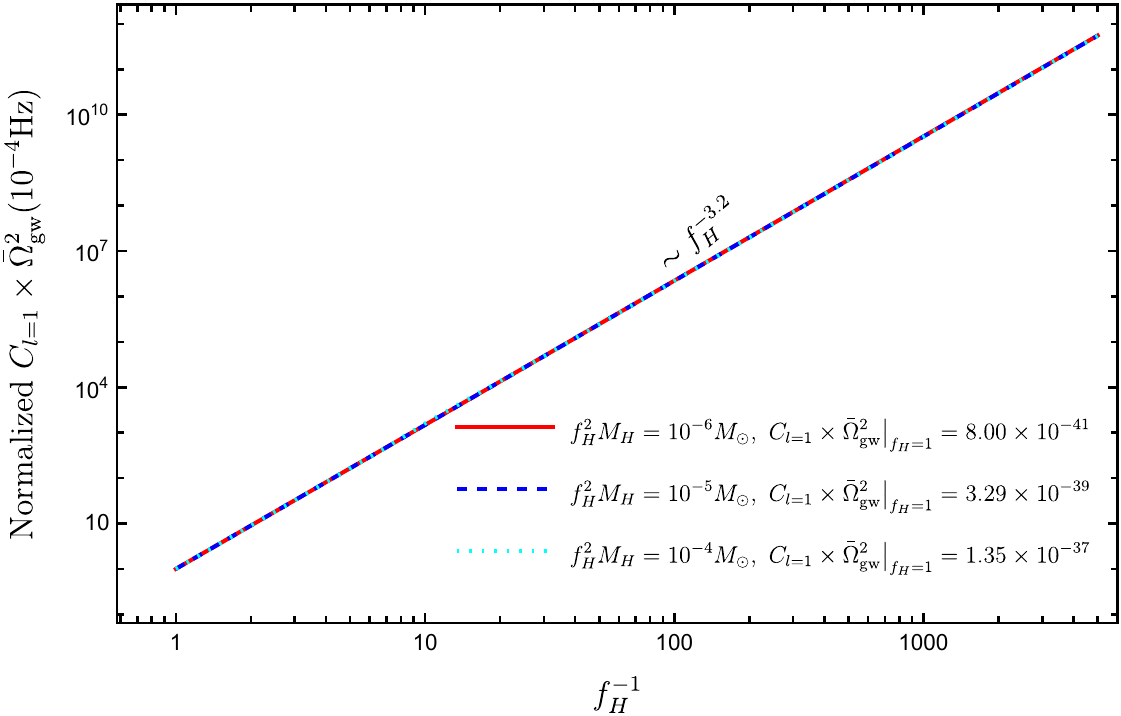}\qquad
  \includegraphics[width=0.45\textwidth]{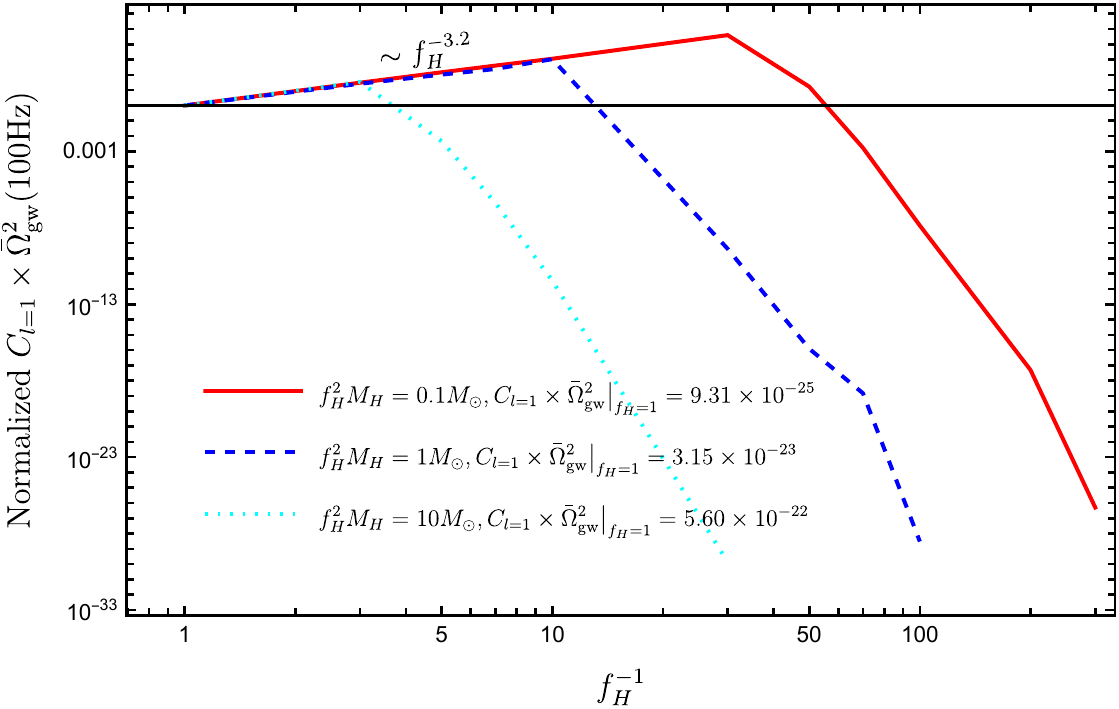}
 \caption{\label{fig:Abs-Cl-ratio-largen} The absolute value of angular power spectra as a function of $f_H^{-1}$ at $10^{-4}$Hz (left) and 100Hz (right) for $l=1$ and different values of $f_H^2M_H$ normalized by the corresponding $C_{l=1}\times\bar{\Omega}_{\rm gw}^2\big\vert_{f_H=1}$, with ${f_H}_{\rm min}=5000^{-1}$ (left) and $f_{H,{\rm min}}=300^{-1}$ (right). The black solid line in the right panel indicates the value of 1.}
\end{figure}

Even though $\sigma_0^2$ is highly suppressed at a large $f_H^{-1}\gtrsim10$ for a given $A_s$, we further study the regime where $f_H^{-1}\gg 1$ for completeness, noting that this regime is also theoretically motivated by several deeply sub-horizon formation mechanisms \cite{Cotner:2019ykd,PhysRevD.26.2681}. We note that since we require $f_{\rm PBH}=0.01$ in the plots, the value of $\sigma_0^2$ is almost unchanged even for large $f_H^{-1}$ by manually tuning $A_s$. Figure \ref{fig:Abs-Cl-ratio-largen} (left) shows the absolute angular power spectra $C_l\times\bar{\Omega}_{\rm gw}$ of $l=1$ and $f_{\rm PBH}=0.05$ as a function of $f_H^{-1}$ at $\nu=10^{-4}{\rm Hz}$ and $f_H^2M_{H}\in\{10^{-6}M_\odot,10^{-5}M_\odot,10^{-4}M_\odot\}$, while Figure \ref{fig:Abs-Cl-ratio-largen} (right) shows the similar angular power spectra at $\nu=100{\rm Hz}$ and $f_H^2M_{H}\in\{0.1M_\odot,1M_\odot,10M_\odot\}$. There is a manifest power-law behavior for the left subplot across the whole range of $f_H^{-1}$ and for the right subplot right before the spectrum drops significantly due to the fact that the observed frequency is far above the cutoff frequency of the GW energy spectrum.  


The $f_H$-dependence or effectively $M_{\rm PBH}$-dependence of the absolute angular power spectrum can be understood as follows: First, when considering the observed frequency in the inspiral phase of the IMR spectrum and the two PBHs in the binary have the same mass, we notice the $f_H$-dependence of the isotropic SGWB energy density :
\be
\bar{\Omega}_{\rm gw,inspiral}\propto \mathcal{L}\propto \frac{dE}{d\nu}\bigg\vert_{\rm inspiral}\times\mathcal{R}\propto f_H^{-1.6},
\ee
where the power law index, $1.6\approx\left(\frac{5}{3}-\frac{32}{37}\right)\times 2$, is determined by the binary PBH mass dependence from both the merger rate and GW energy spectrum,
i.e., $\mathcal{R}\propto M_b^{-32/37}$ and $\frac{dE}{d\nu}\bigg\vert_{\rm inspiral}\propto M_b^{5/3}$. By using the above power-law $f_H$-dependence, we can further show that the $C_l\times\bar{\Omega}_{\rm gw}^2$ ratio (shown in Figure \ref{fig:Abs-Cl-ratio-largen}) has the power-law behavior with respect to $f_H$:
\be
C_l\times\bar{\Omega}_{\rm gw}^2(f_H)\propto \mathcal{L}^2\propto \left( \frac{dE}{d\nu}\bigg\vert_{\rm inspiral}\times\mathcal{R}\right)^2\propto f_H^{-3.2},
\ee
where we have used the fact that $f_H$-dependence on $C_l$ is sub-dominant compared to the one on $\bar{\Omega}_{\rm gw}^2$, which can be found by considering the cancellation of $f_H$-dependence between $\bar{\Omega}_{\rm gw}$ and $\mathcal{L}$ in Eq.~(\ref{eq:clratio}) and can be cross-checked by comparing Figure \ref{fig:Cl-100hz-frac-n} and Figure \ref{fig:Abs-Cl-ratio-largen} for $\nu=100$ Hz and $10^{-4}$Hz. 

\begin{figure}
\centering  
  \includegraphics[width=0.46\textwidth]{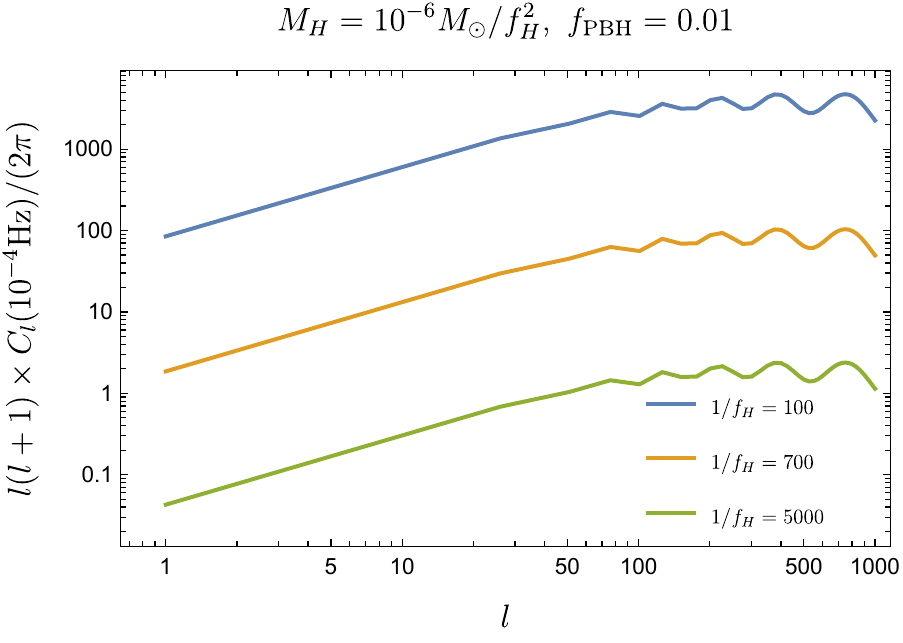}\qquad
  \includegraphics[width=0.48\textwidth]{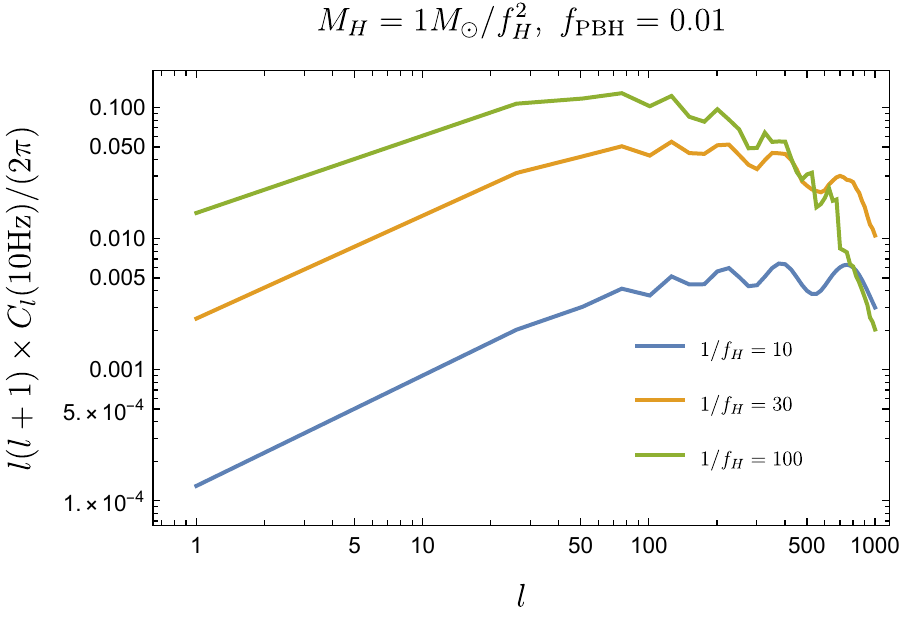}
 \caption{\label{fig:large-n-Cl} The angular power spectra as a function of mutipole $l$ for $f_H^{-1}=100,700,5000$ with $\{\nu=10^{-4}{\rm Hz},~M_{H}=10^{-6}M_\odot\}$ (left) and for $f_H^{-1}=10,30,100$ with $\{\nu=10{\rm Hz},~M_{H}=1M_\odot\}$ (right). }
\end{figure}
Finally, we show in Figure \ref{fig:large-n-Cl} the angular power spectra as a function of $l$ given large values of $f_H^{-1}=100$, $700$, and $5000$ at $\{\nu=10^{-4}{\rm Hz},~M_{H}=10^{-6}M_\odot\}$ (left panel), and the corresponding ones for $f_H^{-1}=10$, $30$, and $100$ at $\{\nu=10{\rm Hz},~M_{H}=1M_\odot\}$ in Figure \ref{fig:large-n-Cl} (right panel). There is no unexpected behavior in the $C_l$ ratio for different $l$ except the case for $f_H^{-1}=100$ in the right subplot. This is, again, because $M_{\rm PBH}$ increases with $f_H^{-2}$, and as a result the cutoff frequency of the IMR waveform shifts to a lower frequency which is far below the observed frequency.

\section{Discussion and Conclusion}\label{sec:discussion}
We performed, for the first time, a comprehensive study of the behavior of the isotropic SGWB energy density spectrum and the angular power spectrum of PBH binary mergers for PBH forming from the collapse of sub-horizon patches. We introduced a model-independent parameter $f_H$ characterizing the fraction of the wavelength of the perturbation mode in units of the horizon radius where the patch gravitationally collapses into a BH. The standard scenario is recovered for $f_H=1$, whereas values $f_H<1$ correspond to sub-horizon collapsing regions. We illustrated our findings using a simple Dirac-$\delta$ peak curvature spectrum, but the spectrum shape can be modified easily depending on the specific sub-horizon PBH formation mechanism. One of the limitations in our results is the assumption of a spherically symmetric distribution of overdensities, which in general can be non-spherically distributed when entering the regime of very small $f_H$.

We found that the sub-horizon PBH formation in general {\em enhances} the isotropic SGWB energy density $\bar{\Omega}_{\rm gw}$ and the absolute angular power spectrum $C_l\times\bar{\Omega}^2_{\rm gw}$. However, the almost monotonic increases in both $\bar{\Omega}_{\rm gw}$ and $C_l\times\bar{\Omega}^2_{\rm gw}$, as $f_H$ decreases, cease when the chirp mass of binary PBHs reaches a mass threshold determined by a given observed frequency, such that the isotropic SGWB energy density spectrum significantly drops above that specific cutoff frequency. The important effect of the sub-horizon formation is changing the PBH mass (distribution) and also the formation time or redshift of PBHs, which in turn affects the GW observables, including both $\bar{\Omega}_{\rm gw}$ and $C_l\times\bar{\Omega}^2_{\rm gw}$.

We investigated the isotropic SGWB energy density spectrum and the angular power spectrum at various frequencies, PBH masses, and horizon size fractions $f_H$ during  PBH formation. When the observed frequency sits at the frequency range of the inspiral phase of the IMR waveform, the isotropic SGWB energy density spectrum and the absolute angular power spectrum at that frequency have $f_H$-dependent power-law behaviors, i.e., $\bar{\Omega}_{\rm gw}\propto f_H^{-1.6}$ and $C_l\times\bar{\Omega}^2_{\rm gw}\propto f_H^{-3.2}$.  

By introducing a sub-horizon formation scenario in the calculation of $\bar{\Omega}_{\rm gw}$ and $C_l\times\bar{\Omega}^2_{\rm gw}$, one can study the rich phenomenology of PBHs formed by non-standard mechanisms across the universe's history and provide a way to potentially test PBH formation mechanisms upon the hopefully forthcoming detection of an SGWB signal.   

\section*{Acknowledgments}

This work is supported in part by the U.S. Department of Energy grant number de-sc0010107 (SP). F.W.Y. is supported by the U.S. Department of Energy under Award No. DESC0009959.

\bibliography{ref}
\end{document}